\documentclass[pra,twocolumn,showpacs,showkeys,floatfix,amsmath,amssymb]{revtex4-1}
\usepackage{graphicx,exscale,bm,color}
\usepackage{hyperref}

\begin{document}

\title{Quantum machine learning with glow for episodic tasks and decision games}
\date{\today}
\author{Jens Clausen}
\affiliation{Institut f\"ur Theoretische Physik,
             Universit\"at Innsbruck,
             Technikerstra\ss{}e 21a,
             A-6020 Innsbruck,
	     Austria}
\author{Hans J. Briegel}
\affiliation{Institut f\"ur Theoretische Physik,
             Universit\"at Innsbruck,
             Technikerstra\ss{}e 21a,
             A-6020 Innsbruck,
	     Austria}
\begin{abstract}
We consider a general class of models, where a reinforcement learning (RL) agent
learns from cyclic interactions with an external environment via classical
signals. Perceptual inputs are encoded as quantum states, which are subsequently
transformed by a quantum channel representing the agent's memory, while the
outcomes of measurements performed at the channel's output determine the agent's
actions. The learning takes place via stepwise modifications of the channel
properties. They are described by an update rule that is inspired by the
projective simulation (PS) model and equipped with a glow mechanism that allows
for a backpropagation of policy changes, analogous to the eligibility traces in
RL and edge glow in PS. In this way, the model combines features of PS with the
ability for generalization, offered by its physical embodiment as a quantum
system. We apply the agent to various setups of an invasion game and a grid
world, which serve as elementary model tasks allowing a direct comparison with a
basic classical PS agent.
\end{abstract}
\maketitle
\section{
\label{sec1}
Introduction}
\subsection{
Motivation}
If we consider the development of new technologies as a collective learning
process, we can distinguish between different interlaced processes. While basic
research focuses on exploration characterised by a search for potential
alternatives to established methods, the more promising an approach appears, the
more likely it becomes subject to subsequent exploitation, where it is optimised
and matured with the ultimate hope to supersede what is available. An example of
explorative activity are early efforts to solve problems by artificial
intelligence (AI), such as inventing unconventional heuristic techniques. AI has
recently regained interest \cite{NatureSpecialIssue,ScienceSpecialIssue}, which
may be a consequence of new approaches to computation
\cite{bookNielsen,bookWittek} as well as improved capacities of classical
computing and networking. The present work aims at drawing a connection between
the recently suggested scheme of PS \cite{Bri12,Mau15} and quantum control
theory \cite{bookAlessandro,bookWiseman}, restricting attention to example
problems analogous to those considered in the basic classical PS schemes
\cite{Bri12,Mau15,Mel14,Mel15}, rather than a treatment of practically relevant
but large-scale applications of, e.g., machine learning. A discussion of
scalability, quantum speed up, or practical implementability
\cite{Pap14,Dun15,Fri15} is beyond the scope of this work.

We consider a class of schemes, where a quantum agent learns from cyclic
interactions with an external environment via classical signals. The learning
can be considered as an \emph{internal} quantum navigation process of
the agent's ``hardware'' or ``substrate'' that forms its memory of past
experience. For notational convenience, we describe the memory operation as a
unitary $\hat{U}$ involving (information carrying and other) controllable and
uncontrollable degrees of freedom (such as a ``bath''), where the latter are not
necessarily identical with the environment, on which the agent operates. While
conceptually, the memory may hence be seen as an open quantum system
\cite{bookBreuer}, the numerical examples considered in the present work
restrict to closed system dynamics. This navigation of agent memory $\hat{U}$
must be distinguished from the evolution of quantum states in which, following
external or internal stimulus, the memory is excited \cite{Bri12,Pap14}.
Learning as an internal navigation process corresponds to the colloquial notion
of a learner desiring to quickly ``make progress'' rather than ``marking time''.
For the agent's internal dynamics, we talk of a navigation process rather than
a navigation problem that is to be solved, since ultimately, the agent responds
to its environment that is generally unknown and subject to unpredictable
changes.

While the proposed PS-model is characterised by an episodic \& compositional
memory (ECM), we here ignore the clip network aspect and restrict attention to a
parameter updating that is motivated from the basic scheme \cite{Bri12,Mau15},
which we apply to simple learning tasks involving an agent equipped with a
quantum memory. We specifically reconsider some of the examples discussed in
\cite{Bri12,Mel14,Mel15} in order to investigate to what extent the results can
be reproduced. In contrast to the classical scheme, where the parameters are
weights in a clip network, we here refrain from ascribing a particular role,
they could play (e.g., in a quantum walk picture mentioned in \cite{Bri12}).
Here, the parameters are simply controls, although in our examples, they are
defined as interaction strengths in a stack of layers constituting the agent
memory $\hat{U}$. This choice of construction is however not essential for the
main principle. From the viewpoint of the network-based classical PS, drawing a
connection to quantum control theory opens the possibility to apply results
obtained in the latter field over the last years
\cite{Dong06a,*Dong06b,*Dong08a,*Dong08b,*Dong09,*Dong10,*Dong14,*Dong15}.
On the other
hand, classical PS is similar to RL \cite{bookSuttonBarto,bookRussellNorvig},
which considers a type of problems, where an ``agent'' (embodied decision maker
or ``controller'') learns from interaction with an environment (controlled
system or ``plant'') to achieve a goal. The learning consists in developing a
(generally stochastic) rule, the agent's ``policy'', of how to act depending on
the situation it faces, with the goal to accumulate ``reward'' granted by the
environment. In RL, the environment is anything outside of control of this
decision making. The reward could describe for example pleasure or pain felt by
an individual. It is generated within the individual's body but is beyond it's
control, and therefore considered originating in the \emph{agent's} environment.
 Historically, RL, which must be distinguished from supervised
learning, originates from merging a trait in animal psychology with a trait in
control theory. Although dynamic programming as the basis of the latter is well
understood, limited knowledge of the environment along with a vast number of
conceivable situations, an RL-agent may face, render a direct solution
impossible in practice. Analogous to RL growing out of dynamic programming by
refining the updates of values, in a quantum context, one could think of
refining quantum control schemes with algorithmic elements that enhance their
resource efficiency.

Another aspect is embodiment. A historical example is application-specific
classical optical computing with a 4F-optical correlator. A more recent effort
is neuromorphic computing, which aims at a very-large-scale integration
(VLSI)-based physical implementation of neural networks, whose simulation with
a conventional computer architecture is inefficient. This becomes even more
crucial for quantum systems, which may be implemented as superconducting solid
state devices, trapped ions or atoms, or wave guide-confined optical fields.
Given the availability of a controllable quantum system, it is hence tempting to
transform quantum state-encoded sensory input and select actions based on
measurement outcomes. While the parameter update is typically done by some
standard linear temporal difference (TD)-rule, the selection of actions is in
classical algorithms governed by a separate stochastic rule that tries to
balance exploration and exploitation. This stochastic rule is described in terms
of a policy function, that determines, how the probabilities for choosing the
respective actions depend on the value functions in RL, edge strengths in PS, or
controls in direct policy approaches. Examples are the $\varepsilon$-greedy and
the softmax-rule. The quantum measurement here serves as a direct
physical realisation of an action-selection, whose uncertainty allows to
incorporate both exploration and exploitation \cite{Dong08b}. In our context,
the resulting measurement-based (and hence effectively quadratic) policy forms
an intermediate between the linear stochastic function used in \cite{Bri12} and
the exponential softmax-function applied in \cite{Mel14}. A measurement-based
policy can moreover be tailored on demand by the way in which classical input is
encoded as a quantum state. One could, e.g., apply mixtures of a pure state and
a maximally mixed state to mimic an $\varepsilon$-greedy policy function, or one
could use thermal input states to mimic an exponential function. In contrast to
the value function-based RL, our approach amounts to a direct policy search,
where the agent-environment interaction employs a general state preparation
$\to$ transformation $\to$ measurement scheme, that reflects the kinematic
structure of quantum mechanics.
\subsection{
\label{sec3}
RL as a reward-driven navigation of the agent memory}
Consider the specific task of mapping input states $|s_i\rangle$ by means of a
controllable unitary $\hat{U}$ to outputs $|a_i\rangle$. Under the (restrictive)
assumption, that for each input there is exactly one correct output, the task is
to learn this output from interaction with an environment. In our context, the
$|s_i\rangle$ ($|a_i\rangle$) are regarded as encoded percepts (actions), while
$\hat{U}$ acts as memory of the learned information and can finally accomplish
the mapping as an embodied ``ad hoc'' computer or an ``oracle'', which is
similar to learning an unknown unitary \cite{Ban14}.

Consider (i) the case where there is only one possible input state $|s\rangle$.
If the task is the navigation of the output state
$\hat{\varrho}$ $\!=$ $\!\hat{U}|s\rangle\langle{s}|\hat{U}^\dagger$ by means of
$\hat{U}$ to a desired destination state $|a\rangle$, a learning agent has to
realize the maximisation of the conditional probability
$p(a|s)$ $\!=$ $\!\langle{a}|\hat{\varrho}|a\rangle$ by tuning $\hat{U}$. The
intuition behind this is that $p$ is bounded and if $\hat{U}(\bm{h})$ depends
analytically on some control vector $\bm{h}$, the gradient with respect to
$\bm{h}$ must vanish at the maximum of $p$. To give a simple example, we assume
that $\hat{U}(t)$ depends (rather than on $\bm{h}$) on a single real parameter
$t$ in a continuous and differentiable way such that it obeys the
Schr{\"o}dinger equation {\small $\frac{\mathrm{d}}{\mathrm{d}t}\hat{U}$
$\!=$ $\!-\mathrm{i}\hat{H}\hat{U}$} with the state boundary conditions 
$\hat{\varrho}(0)$ $\!=$ $\!|s\rangle\langle{s}|$ and
$\hat{\varrho}(t_{\mathrm{F}})$ $\!=$ $\!|a\rangle\langle{a}|$. This gives 
\begin{eqnarray}
  \frac{\mathrm{d}p(t)}{\mathrm{d}t}&=&2\mathrm{Re}\langle{a}|
  \Bigl(\frac{\mathrm{d}}{\mathrm{d}t}\hat{U}\Bigr)|s\rangle\langle{s}|
  \hat{U}^\dagger|a\rangle
  \\
  &=&2\mathrm{Im}\langle{a}|\hat{H}\hat{\varrho}(t)|a\rangle,
\end{eqnarray}
so that indeed
\begin{equation}
  \frac{\mathrm{d}p(t)}{\mathrm{d}t}|_{t=t_{\mathrm{F}}}
  =2\mathrm{Im}\langle{a}|\hat{H}|a\rangle=0.
\end{equation}
Any algorithm that results in a $\hat{U}$ such that $p$ approaches 1
accomplishes this task.

Assume now that (ii) we are required to transform a given orthonormal basis
(ONB) $\{|s_i\rangle\}$ into another given ONB $\{|a_i\rangle\}$ of a vector
space of same dimension, but we are not told which state is to be transformed
into which other state. We could build a quantum device that implements some
unitary $\hat{U}_{\mathrm{T}}$ such that
$|a_i\rangle$ $\!=$ $\!\hat{U}_{\mathrm{T}}|s_i\rangle$. Preparing the system in
state $|s_i\rangle$ and measuring in the second basis gives outcome
$|a_i\rangle$. One may consider the problem as a (trivial) learning task, namely
that of an identical mapping of the state-indices $i$. However, if we do not
know from the beginning what kind of mapping the solution is, we have to learn
it. In our quantum device, we would tune $\hat{U}$ until it gives the desired
measurement statistics. Inspired by \cite{Bri12}, we call this task
``invasion game''. To solve it, we initialize the device in states $|s_i\rangle$
chosen randomly from the given ONB, while the measurement is done in the second
ONB formed by the $|a_i\rangle$. The algorithm will drive $\hat{U}$ to some
unitary $\hat{U}_2^\prime\hat{U}_{\mathrm{T}}\hat{U}_1$, where
$\hat{U}_1$ ($\hat{U}_2^\prime$) are undetermined unitaries which are diagonal
in the basis $\{|s_i\rangle\}$ ($\{|a_i\rangle\}$).

If (iii) the percept states are random, this phase freedom is removed up to a
global phase. In the simplest case, we draw the initial states of the device
from an ``overcomplete'' basis, where the set of all possible states is linearly
dependent. For a $n$-level system, this can be accomplished by (randomly)
choosing $n$ SU($n$)-unitaries. During each state initialisation, we then take
one $\hat{U}_{\mathrm{R}}$ from this set, a random $|s_i\rangle$ from our first
ONB, and then prepare the device in a state $\hat{U}_{\mathrm{R}}|s_i\rangle$.
Consequently, the measurement is done in a transformed basis formed by the
$\hat{U}_{\mathrm{T}}\hat{U}_{\mathrm{R}}\hat{U}_{\mathrm{T}}^{-1}|a_i\rangle$
rather than the $|a_i\rangle$ themselves.

In this sense, the navigation of (i) a given input state, (ii) a given ONB, and
(iii) random states can be described as a navigation of unitaries $\hat{U}$ with
a varying amount of freedom.
While formally, all three cases (i)-(iii) can be considered as special cases of
a navigation of $\hat{U}(\bm{h})$ to a point (\ref{task}), where a percept
statistics-based fidelity (\ref{F2}) becomes maximum, practically they can be
accomplished in RL by means of the mentioned reward signal, independently of the
availability of analytic solutions. In what follows, we consider $\hat{U}$ as a
memory of an RL-agent, that solves tasks arising from its interaction with an
environment.
\section{
\label{sec2}
A cybernetic perspective}
\begin{figure}[ht]
\includegraphics[width=6cm]{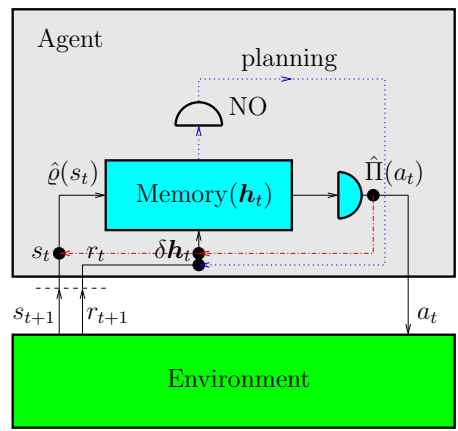}
\caption{\label{fig1}
Agent-environment interaction as a feedback scheme. The $s$ are percepts, which
initialise the agent's memory in a quantum state $\hat{\varrho}({s})$. Choice of
an action $a$ is made by a measurement process as described by a given POVM
$\hat{\Pi}$. Depending on the rewards $r$ given by the environment, the memory
is updated at the end of each cycle $t$. The memory can also be modified by
internal loops based on a numerical objective NO (dotted line) or measurements
(dash-dotted line).
}
\end{figure}

\begin{figure}[ht]
\includegraphics[width=6cm]{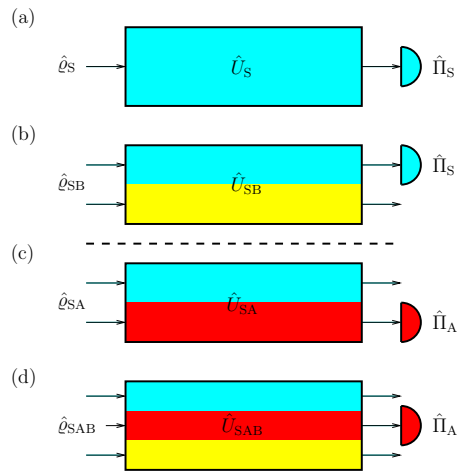}
\caption{\label{fig2}
Examples for the agent's memory as shown in Fig.~\ref{fig1}. (a) unitary
evolution of the percept states, (b) open system evolution due to interaction
with a bath B, (c) composite memory with coupled subsystems for percept (S) and
action (A) variables, (d) extends (c) to an open system evolution analogous to
(b) extending (a). Further ancilla systems may be added (not shown), to account
for, e.g., emotion degrees of freedom introduced in \cite{Bri12}.
}
\end{figure}

The scheme is depicted in Fig.~\ref{fig1}.
The agent is equipped with some quantum channel that acts as its memory whose
properties can be modified by control parameters denoted by a vector $\bm{h}$.
Examples of memory structures are listed in Fig.~\ref{fig2}.
In Fig.~\ref{fig2} and in what follows, we refer to the memory operation by
means of some unitary $\hat{U}$ for notational simplicity. Since any quantum
process can be treated as unitary on an enlarged space, this is not a conceptual
restriction. The agent interacts with an external environment in discrete
cycles $t$. At the beginning of a cycle, the agent receives (via sensors) some
percept $s$, which it encodes as a quantum state $\hat{\varrho}({s})$, in which
its memory is prepared. After transformation of $\hat{\varrho}({s})$ by the
memory channel, a quantum measurement is performed, where we assume for
simplicity that the positive operator valued measure (POVM) $\{\hat{\Pi}\}$
describing this measurement is fixed. Depending on the outcome of this
measurement, an action $a$ is selected and performed on the environment (via
actuators), which completes the cycle. The environment reacts with a new percept
and a reward $r$, which are perceived by the agent during the following cycle.
Depending on the reward, some adjustments are made on the control parameters,
which modify the properties of the memory channel (i.e., its ``hardware''). This
feedback loop is adapted from the classical schemes in \cite{bookSuttonBarto}
and \cite{bookRussellNorvig}, where the percepts $s$  in Fig.~\ref{fig1}
correspond to the states in \cite{bookSuttonBarto}. The agent's interaction with
the environment is here considered classical in the sense that percepts, actions
and rewards are classical signals. The environment itself is not specified, it
could represent, e.g., an experiment performed on a quantum system. Note that
the environment in Fig.~\ref{fig1} is not to be confused with the bath in
Fig.~\ref{fig2}, which affects the memory channel but is not considered part of
the agents ``habitat''.

In addition to the external loop, we may also equip the agent with two types of
internal feedback loops, which allow the agent to undertake what corresponds to
``planning steps'' in RL and ``reflection'' in PS. One type is similar to the
external loop in that it involves state initialisations and measurements on the
memory channel, but exploits that percepts, actions and rewards can be recorded
and reproduced as a consequence of their classicality. The second type of
internal loop does not involve state evolutions but requires some mathematical
model of the memory channel itself, which is used to directly calculate a
numerical objective (NO), whose value is used to alter the control parameters.
Fig.~\ref{fig1} does not imply that all of these loops need to be
simultaneously present, they are rather thought of either subprocesses within an
overall agent scheme or possible modes of its operation. The numerical examples
in this work will exclusively apply the external loop.

All three loops involve a parameter update $\delta\bm{h}$. In a ``first-order''
update, $\delta\bm{h}$ is proportional to some quantity that depends on the
gradient $\bm{\nabla}\hat{U}$ of $\hat{U}$ with respect to $\bm{h}$. This
gradient can either be computed directly from a memory model $\hat{U}(\bm{h})$
(i.e., from some symbolic expression of $\bm{\nabla}\hat{U}$ if available) or
estimated from measurements. These ``measurements'' can be physical (POVM in
Fig.~\ref{fig1}) or numerical (NO in Fig.~\ref{fig1}). For the estimation, one
varies the components of $\bm{h}$ by a small amount and records the changes in
the measured POVM or computed NO. Here are some elementary examples:
\emph{(1a)} A simulation of an external loop with a given model-based
(i.e. analytic) $\bm{\nabla}\hat{U}$ is performed in Sec.~\ref{sec:ig22}
(Fig.~\ref{fig5}) for the case Fig.~\ref{fig2}(c), in Sec.~\ref{sec:ig44}
(Figs.~\ref{fig7}-\ref{fig8}) for the case Fig.~\ref{fig2}(a), and in
Sec.~\ref{sec:gw} (Figs.~\ref{fig11} and \ref{fig12}) for the case
Fig.~\ref{fig2}(c). 
\emph{(1b)}  A simulation of an external loop with a POVM measurement-based
$\bm{\nabla}\hat{U}$ is carried out in \cite{Cla15} (Fig. 6) for the case
Fig.~\ref{fig2}(b).
\emph{(2)} A NO-based internal loop with a model-based $\bm{\nabla}\hat{U}$ is
considered in \cite{clausen18} for the case Fig.~\ref{fig2}(b) and in
\cite{Cla15} (Figs.2-4) for the case Fig.~\ref{fig2}(a).
\emph{(3)} The POVM-based internal loop in Fig.~\ref{fig1} can be used to
estimate $\bm{\nabla}\hat{U}$ in the absence of a model $\hat{U}(\bm{h})$ of the
agent memory. To this end, one of the agent's possibilities consists in
inserting a number of internal cycles between each external cycle, where it
repeatedly prepares its memory in the latest percept state and observes how a
variation $\delta\bm{h}$ affects the measurement statistics. A discussion of
this will be given in Sec.~\ref{sec6}. Beyond these examples, all three loops
can be interlaced with each other in various ways, analogous to the wealth of
approaches reviewed in \cite{bookSuttonBarto}.
\section{Update rule in parameter space}
For the cycle-wise update of the control parameters $\bm{h}$ of the memory
channel $\hat{U}$, we apply a rule
\begin{equation}
\label{ur3}
  \bm{h}^\prime
  =\bm{h}+\kappa(\bm{h}_\infty-\bm{h})
  +{\alpha}{r}\sum_{k=0}^{t-1}(1-\eta)^k\bm{D}_{t-k},
\end{equation}
inspired by the basic model of PS \cite{Bri12,Mau15}. The number of components
$h_k$ can range from one ($\bm{h}$ scalar) to infinity ($\bm{h}$ may represent a
function or a vector of functions), and the $h_k$ can be assumed to be
real-valued without loss of generality. In \cite{Bri12,Mau15}, the components of
$\bm{h}$ are the edge strengths of a directed graph representing a network of
clips (the graph's vertices). While these clips are considered sequences of
remembered percepts and actions, the network itself abstracts from the clip's
internal contents. Our view of $\bm{h}$ as a control vector is one further
simplification and generalization that may allow for but does not require the
view of the memory as a network.

In (\ref{ur3}), $\bm{h}$ and $\bm{h}^\prime$ are the control vectors before and
after the update at cycle $t$, respectively. $\alpha$ $\!\ge$ $\!0$ is a
(typically small) learning rate, and $r$ is the reward given at cycle $t$.
$\kappa$ $\!\in$ $\![0,1]$ is a relaxation rate towards some equilibrium value
$\bm{h}_\infty$ in the absence of rewards. This allows for what corresponds to
the ``forgetting'' process suggested in \cite{Bri12,Mau15} to account for
dissipation in an embodied implementation and deal with time-dependent
environments. A natural possibility is to identify the value $\bm{h}_0$, with
which the memory is initialised before the first cycle, with $\bm{h}_\infty$.
This could be the zero vector $\bm{h}_0$ $\!=$ $\!\bm{0}$ yielding, e.g., the
identity, $\hat{U}_0$ $\!=$ $\!\hat{U}(\bm{h}_0)$ $\!=$ $\!\hat{I}$. The
learning process will then be a reward-driven and generally stochastic
navigation in parameter space $\{\bm{h}\}$ away from the zero vector $\bm{0}$.
Lifted to $\hat{U}(\bm{h})$, this navigation starts at the identity
$\hat{U}_0$ $\!=$ $\!\hat{I}$, that relaxes back to it in the prolonged absence
of rewards. In this work, we consider static environments as in \cite{Mel14},
and hence always set $\kappa$ $\!=$ $\!0$. $\bm{D}_{t}$ is a difference vector.
While some options for finite difference choices of $\bm{D}$ are outlined in
Sec.~\ref{sec6}, in all numerical examples within this work we restrict to the
case, where $\bm{D}_{t}$ $\!=$ $\!\bm{\nabla}_{t}$ is a short-hand notation for
the gradient
\begin{eqnarray}
\label{grad}
  \bm{\nabla}_{t}&=&\bm{\nabla}p({a}|{s})_t=2\mathrm{Re}
  \bigl\langle\hat{U}^\dagger\hat{\Pi}({a})\bm{\nabla}\hat{U}\bigr\rangle_{t},
\\
\label{grada}
  p({a}|{s})&=&\mathrm{Tr}\bigl[\hat{U}\hat{\varrho}({s})\hat{U}^\dagger
  \hat{\Pi}({a})\bigr]
  =\bigl\langle\hat{U}^\dagger\hat{\Pi}({a})\hat{U}\bigr\rangle,
\end{eqnarray}
with components $\frac{\partial{p}({a}|{s})}{\partial{h}_k}$ at cycle $t$.
$p({a}|{s})$ is the probability of the obtained measurement outcome ${a}$ under
the condition of the respective cycle's percept state $\hat{\varrho}({s})$,
where
$\langle\cdots\rangle$ $\!\equiv$ $\!\mathrm{Tr}[\hat{\varrho}({s})\cdots]$
denotes the expectation value with respect to this state, and $\hat{\Pi}({a})$
is the member of the POVM that corresponds to measurement outcome $a$. The
latter determines the action performed by the agent, and we use the same symbol
for both. $(1$ $\!-$ $\!\eta)$ describes a backward-discount rate, which we have
defined via a parameter $\eta$ $\!\in$ $\![0,1]$ to allow comparison with the
glow mechanism introduced in \cite{Mau15}.
As mentioned above, the unitary transformation of the respective percept states
$\hat{\varrho}({s}_t)$ by the memory $\hat{U}$ $\!=$ $\!\hat{U}(\bm{h}_t)$ at
cycle $t$ in (\ref{grada}) refers in general to a larger (dilated) space.
The dynamical semigroup of CPT maps proposed in \cite{Bri12} is included and
can be recovered by referring to Fig.~\ref{fig2}(d) [or alternatively
Fig.~\ref{fig2}(b)] and the assumption that
\begin{eqnarray}
  \hat{\varrho}({s}_t)\quad(\equiv\hat{\varrho}_{\mathrm{SAB}})
  &=&\hat{\varrho}_{\mathrm{SA}}({s}_t)
  \otimes\hat{\varrho}_{\mathrm{B}},
\\
  \mathrm{Tr}_{\mathrm{B}}\left[\hat{U}\hat{\varrho}({s}_t)
  \hat{U}^\dagger\right]&=&\mathrm{e}^{\mathcal{L}_{\mathrm{SA}}
  \Delta{T}^{(\mathrm{mem})}_t}\hat{\varrho}_{\mathrm{SA}}({s}_t),
\end{eqnarray}
where the physical memory evolution time $\Delta{T}^{(\mathrm{mem})}_t$ may
depend on the cycle $t$ for a chosen parametrisation $\hat{U}(\bm{h})$ and must
be distinguished from the agent response time that can additionally be affected
by the potential involvement of internal loops in Fig.~\ref{fig1}. The
superoperator $\mathcal{L}$ $\!=$ $\!\mathcal{L}_{\mathrm{SA}}$, whose effect on
$\hat{\varrho}$ $\!=$ $\!\hat{\varrho}_{\mathrm{SA}}$ is defined as a sum
\begin{equation}
\label{Lindblad}
  \mathcal{L}\hat{\varrho}=-\mathrm{i}[\hat{H},\hat{\varrho}]+L\hat{\varrho},
\end{equation}
generates in \cite{Bri12} a quantum walk and is given by a Hamiltonian
$\hat{H}$ $\!=$ $\!\sum_{\{j,k\}\in{E}}\lambda_{jk}(\hat{c}_{kj}$ $\!+$
$\!\hat{c}_{kj}^\dagger)$ $\!+$ $\!\sum_{j\in{V}}\epsilon_{j}\hat{c}_{jj}$
and a Lindbladian $L\hat{\varrho}$ $\!=$ 
$\!\sum_{\{j,k\}\in{E}}\kappa_{jk}(\hat{c}_{kj}\hat{\varrho}
\hat{c}_{kj}^\dagger$ $\!-$ $\!\frac{1}{2}
\{\hat{c}_{kj}^\dagger\hat{c}_{kj},\hat{\varrho}\})$, with
$\hat{c}_{kj}$ $\!=$ $\!|{c}_{k}\rangle\langle{c}_{j}|$ performing transitions
between clip states $|{c}_{l}\rangle$ $\!\in$ $\!\mathcal{H}_{\mathrm{SA}}$
along a graph $G$ $\!=$ $\!(V,E)$ consisting of a set $V$ of vertices and a set
$E$ of edges. Since on the one hand, we here do not intend to necessarily
represent $\hat{U}$ by a clip network, and on the other hand do not want to
exclude from the outset situations involving time-dependent or non-Markovian
bath effects \cite{DSH14}, we use the dilated $\hat{U}$ for simplicity instead.
The set of all probabilities $p({a}|{s})$ in (\ref{grada}), i.e, the whole
conditional distribution then defines the agent's policy.

A reward given by the environment at time $t$ raises the
question of the extent to which decisions made by the agent in the past have
contributed to this respective reward. A heuristic method is to attribute all
past decisions, but to a lesser degree the further the decision lies in the
past. (Considering the agent's life as a trajectory of subsequent percepts and
actions, we could imagine the latest event trailing a decaying tail behind.)
A detailed description of this idea is presented in \cite{bookSuttonBarto} in
form of the eligibility traces, which can be implemented as accumulating or
replacing traces. In the context of PS, a similar idea has been introduced as
glow mechanism that can be implemented as edge or clip glow \cite{Mau15,Mel14}.
In our context (\ref{ur3}), we realise it by updating the control vector by a
backward-discounted sum of gradients $\bm{\nabla}_{t-k}$ referring to percepts
and actions involved in cycles that happened $k$ steps in the past. A sole
update by the present gradient is included as limit $\eta$ $\!=$ $\!1$, for
which (\ref{ur3}) reduces to
$\bm{h}^\prime$ $\!=$ $\!\bm{h}+\alpha{r}\bm{\nabla}p({a}|{s})$.
This special case is sufficient for the invasion game, which we will consider
in Sec.~\ref{sec4}, because at each cycle, the environment provides a feedback
on the correctness of the agent's decision by means of a non-zero reward. After
that, we apply the general update (\ref{ur3}) to a grid world task, where the
agent's goal cannot be achieved by a single action, and where the long term
consequences of its individual decisions cannot be foreseen by the agent. 
\section{Relation to existing methods}
In this section, we ignore the embodied implementation of our method as a
quantum agent and briefly summarise and compare the update rules of the methods
considered from a computational point of view. It should be stressed that RL is
an umbrella term for problems that can be described as agent-environment
interactions characterised by percepts/states, actions, and rewards. Hence
\emph{all} methods considered here are approaches to RL-problems. For notational
convenience however, we here denote the standard value function-based methods
as ``RL'' in a closer sense, keeping in mind that alternatives such as direct
policy search deal with the same type of problem. The standard RL-methods
successively approximate for each state or state action pair the expected return
$R_t$ $\!=$ $\!\sum_{k=0}^\infty\gamma^kr_{t+k+1}$, i.e., a sum of future
rewards $r$, forward-discounted by a discount rate $\gamma\in[0,1]$, that the
agent is trying to maximise by policy learning. Corrections to the current
estimates can be done by shifting them a bit towards actual rewards observed
during an arbitrarily given number $n$ of future cycles, giving rise to
``corrected $n$-step truncated returns''
$R_t^{(n)}$ $\!=$ $\!{r}_{t+1}$ $\!+$ $\!\gamma{r}_{t+2}$ $\!+$ $\!\ldots$ $\!+$
$\!\gamma^{n-1}{r}_{t+n}$ $\!+$ $\!\gamma^{n}V_t(s_{t+n})$, where $V$ is the
value function of the respective future state $s_{t+n}$ (analogous
considerations hold for state action pairs). A weighted average of these gives
the $\lambda$-return
$R_t^{\lambda}$ $\!=$ $\!(1-\lambda)\sum_{n=1}^\infty\lambda^{n-1}R_t^{(n)}$,
where $\lambda\in[0,1]$ is a parameter. In an equivalent ``mechanistic''
backward view, this gives rise to so-called eligibility traces. Since the glow
mechanism of PS is closely related to this, we base our comparison on the
$\lambda$-extension of one-step RL. $\lambda$ $\!=$ $\!0$ describes the limit of
shallow sample backups of single-step learning, whereas the other limit
$\lambda$ $\!=$ $\!1$ refers to the deep backups of Monte Carlo sampling, cf.
Fig. 10.1 in \cite{bookSuttonBarto}.

It would be a futile task to try a mapping of the numerous variations,
extensions, or combinations with other approaches that have been discussed or
are currently developed for the methods mentioned, such as actor-critic methods
or planning in RL, or emotion, reflection, composition, generalization, or
meta-learning in PS. In particular, our notion of basic PS implies a restriction
to clips of length $L$ $\!=$ $\!1$, which reduces the edge strengths in the ECM
clip network Fig. 2 in \cite{Bri12} to values $h(s,a)$ of state-action pairs.
Furthermore, in this section, we restrict attention to the basic versions that
are sufficient to treat the numerical example problems discussed in this work.
We may think of tasks such as grid world, where actions lead to state
transitions, until a terminal state has been reached, which ends the respective
episode, cf. Sec.~\ref{sec4}.
\subsection{Tabular RL}
In tabular RL, the updates are performed according to
\begin{eqnarray}
\label{TRL1}
  {e}&\leftarrow&({e}+)1
  \quad[\mathrm{for}\;s\;\mathrm{or}\;(s,a)\;\mathrm{visited}],
  \\
  U&\leftarrow&U+\alpha\left[r+\gamma{U}^\prime-{U}\right]{e}
  \nonumber\\
\label{TRL2}
  &&=(1-\alpha{e})U+\alpha{e}r+\alpha\gamma{e}{U}^\prime,
  \\
\label{TRL3}
  {e}&\leftarrow&\gamma\lambda{e},
\end{eqnarray}
where ${U}$ $\!=$ $\!V(s)$ is a state value function in TD($\lambda$), cf.
Fig. 7.7 in \cite{bookSuttonBarto}, whereas ${U}$ $\!=$ $\!Q(s,a)$ is an action
value function in SARSA($\lambda$), cf. Fig. 7.11 in \cite{bookSuttonBarto}. 
${U}^\prime$ $\!=$ $\!V(s^\prime)$
[or ${U}^\prime$ $\!=$ $\!Q(s^\prime,a^\prime)$] refers to the value of the
subsequent state or state action pair.
${e}$ $\!=$ $\!e(s)$ [${e}$ $\!=$ $\!e(s,a)$] denote the eligibility trace in
TD($\lambda$) [SARSA($\lambda$)]. They can be updated by accumulating
(${e}$ $\!\leftarrow$ $\!{e}$ $\!+$ $\!1$) or replacing
(${e}$ $\!\leftarrow$ $\!1$) them in (\ref{TRL1}) (ignoring other options such
as clearing traces \cite{bookSuttonBarto}). $\alpha$ is a learning rate, $r$ is
the reward, and $\gamma$ the discount rate. Note that there are alternatives to
(\ref{TRL1})-(\ref{TRL3}). One of them is Q-learning, which can be derived from
SARSA=SARSA($\lambda=0$) by updating $Q(s,a)$ off-policy, which simplifies a
mathematical analysis. Since a Q($\lambda$)-extension of Q-learning is less
straightforward, and there are convergence issues with respect to the
gradient-ascent form discussed below (cf. Sec. 8.5 in \cite{bookSuttonBarto}),
while the methods discussed here update on-policy, we restrict attention to
(\ref{TRL1})-(\ref{TRL3}).
\subsection{Gradient-ascent RL}
Tabular RL is a special case of gradient-ascent RL, where $U$ is in the latter
defined as in (\ref{TRL1})-(\ref{TRL3}), except that it is given by a number of
parameters $\theta_k$, which are combined to a vector $\bm{\theta}$. This
parametrisation can be done arbitrarily. In the linear case, the parameters
could be coefficients of, e.g., some (finite) function expansion, where the
functions represent ``features''. Hence ${U}$ $\!=$ $\!U(\bm{\theta})$, and the
components of the gradient $\bm{\nabla}{U}$ are
$\frac{\partial{U}}{\partial{\theta}_k}$, giving rise to a vector $\bm{e}$ of
eligibility traces. The updates (\ref{TRL1})-(\ref{TRL3}) now generalize to
\begin{eqnarray}
\label{GRL1}
  \bm{e}&\leftarrow&\gamma\lambda\bm{e}+\bm{\nabla}{U},
  \\
\label{GRL2}
  \bm{\theta}&\leftarrow&\bm{\theta}
  +\alpha\left[r+\gamma{U}^\prime-{U}\right]\bm{e},
\end{eqnarray}
cf. Sec. 8 in \cite{bookSuttonBarto}. While the eligibility traces are
initialised with zero, the value functions (by means of their parameters) can be
initialised arbitrarily in tabular and gradient-ascent RL.
\subsection{PS}
Classical PS is a tabular model. By tabular we mean that the percepts and
actions (and ultimately also clips of length $L$ $\!>$ $\!1$ in the ECM) form
discrete (i.e., countable) sets, with the consequence, that the edge strengths
$h$ can be combined to a table (matrix). Let us hence write the updates as
summarised in App.~\ref{app:A1} in a form allowing comparison with
(\ref{TRL1})-(\ref{TRL3}):
\begin{eqnarray}
\label{PS1}
  {g}&\leftarrow&1
  \quad[\mathrm{for}\;(s,a)\;\mathrm{visited}],
  \\
  {h}&\leftarrow&{h}+\lambda{g}+\gamma(h^{\mathrm{eq}}-h)
  \nonumber\\
\label{PS2}
  &&=(1-\gamma){h}+\lambda{g}+\gamma{h}^{\mathrm{eq}},
  \\
\label{PS3}
  {g}&\leftarrow&(1-\eta){g}.
\end{eqnarray}
In (\ref{PS1})-(\ref{PS3}), we intentionally adopted the notation of PS. The
glow parameter ${g}$ in (\ref{PS1})-(\ref{PS3}) corresponds to a replacing trace
${e}$ in (\ref{TRL1})-(\ref{TRL3}), with $(1$ $\!-$ $\!\eta)$ in (\ref{PS3})
corresponding to $\gamma\lambda$ in (\ref{TRL3}), and $\lambda$ in (\ref{PS2})
corresponds to the reward $r$ in (\ref{TRL2}). The discount rate $\gamma$ in
(\ref{TRL1})-(\ref{TRL3}) must not be confused with the dissipation or damping
rate $\gamma\in[0,1]$ in (\ref{PS2}). To avoid confusion, we denote the former
by $\gamma_{\mathrm{disc}}$ and the latter by $\gamma_{\mathrm{damp}}$ for the
remainder of this paragraph. If we disregard the absence of a learning rate in
(\ref{PS1})-(\ref{PS3}) [we may set $\alpha$ $\!=$ $\!1$ in
(\ref{TRL1})-(\ref{TRL3})], we can obtain PS from tabular SARSA($\lambda$) by
replacing the action value function $Q(s,a)$ with the connection weight
$h(s,a)$, and the update of $h$ corresponding to the r.h.s. of (\ref{TRL2}),
\begin{equation}
\label{update1}
  (1-{g}){h}+\lambda{g}+\gamma_{\mathrm{disc}}{g}{h}^\prime
  ={h}+\lambda{g}+(\gamma_{\mathrm{disc}}{h}^\prime-{h}){g},
\end{equation}
with the update of $h$ given by the r.h.s. of (\ref{PS2}),
\begin{equation}
\label{update2}
  (1\!-\!\gamma_{\mathrm{damp}}){h}\!+\!\lambda{g}\!+\!
  \gamma_{\mathrm{damp}}{h}^{\mathrm{eq}}
  \!=\!{h}\!+\!\lambda{g}\!+\!
  \gamma_{\mathrm{damp}}({h}^{\mathrm{eq}}\!-\!{h}).
\end{equation}
In (\ref{TRL1})-(\ref{TRL3}), RL is equipped with forward- and
backward-discounting mechanisms, as becomes apparent in the product
$\gamma_{\mathrm{disc}}\lambda$ in (\ref{TRL3}). Disabling the accumulation of
forward-discounted future rewards (that give rise in RL to the return mentioned
above) by setting $\gamma_{\mathrm{disc}}$ $\!=$ $\!0$ reduces (\ref{update1})
to $(1$ $\!-$ $\!g){h}$ $\!+$ $\!\lambda{g}$, while setting
${h}^{\mathrm{eq}}$ $\!=$ $\!0$ reduces (\ref{update2}) to
$(1$ $\!-$ $\!\gamma_{\mathrm{damp}}){h}$ $\!+$ $\!\lambda{g}$. These
expressions are very similar, except that in PS, the constant
$\gamma_{\mathrm{damp}}$ has taken the place of $g$ in RL, so that
$(1$ $\!-$ $\!\gamma_{\mathrm{damp}})$ determines the range of
backward-discounting in (\ref{happ}). Since in (\ref{happ}), it is the
respective past excitations (glowing rewards) $r_t$ $\!=$ $\!g_t\lambda_t$,
rather than the rewards $\lambda_t$ themselves, which is summed up, damping and
glow play a similar role. On the other hand, the factor $(1$ $\!-$ $\!\eta)$
takes in PS the place of $\gamma_{\mathrm{disc}}\lambda$ in (\ref{TRL3}), which
becomes zero together with $\gamma_{\mathrm{disc}}$, as mentioned.
\subsection{Method presented here}
The update rule (\ref{ur3}) is implemented as
\begin{eqnarray}
\label{GM1}
  \bm{e}&\leftarrow&(1-\eta)\bm{e}+\bm{\nabla}\,p(a|s),
  \\
\label{GM2}
  \bm{h}&\leftarrow&
  \bm{h}+{\alpha}{r}\bm{e}+\kappa(\bm{h}_\infty-\bm{h}),
\end{eqnarray}
which can be obtained from gradient-ascent SARSA($\lambda$) by replacing in
(\ref{GRL1})-(\ref{GRL2}) the action value function $Q(s,a)$ with the
conditional probability $p(a|s)$, renaming $\bm{\theta}$ as $\bm{h}$, replacing
$\gamma\lambda$ in (\ref{GRL1}) with $(1$ $\!-$ $\!\eta)$, and replacing in
(\ref{GRL2}) the term
$\alpha\bigl[\gamma{p}(a^\prime|s^\prime)$ $\!-$ $\!{p}(a|s)\bigr]\bm{e}$ with
$\kappa(\bm{h}_\infty-$ $\!$ $\!\bm{h})$. The latter replacement is similar to
the change from (\ref{update1}) to (\ref{update2}), where $\kappa$ in
(\ref{GM2}) corresponds to $\gamma$ in (\ref{PS2}).
Analogous to the comments following (\ref{update1}) and (\ref{update2}), in the
case $\gamma$ $\!=$ $\!0$, the update corresponding to the r.h.s. of
(\ref{GRL2}) becomes
$(\bm{h}$ $\!-$ $\!\alpha{p}\bm{e})$ $\!+$ $\!\alpha{r}\bm{e}$, whereas for
$\bm{h}_\infty$ $\!=$ $\!0$, the r.h.s. of (\ref{GM2}) reduces to
$(1$ $\!-$ $\!\kappa)\bm{h}$ $\!+$ $\!\alpha{r}\bm{e}$. Similar to the tabular
case, the constant damping rate $\kappa$ has in our method taken the place of
$\alpha{p}\bm{e}$ in gradient-ascent RL.
\subsection{
\label{sec:VE}
Can PS be recovered from our approach?}
\begin{figure}[ht]
\includegraphics[width=7cm]{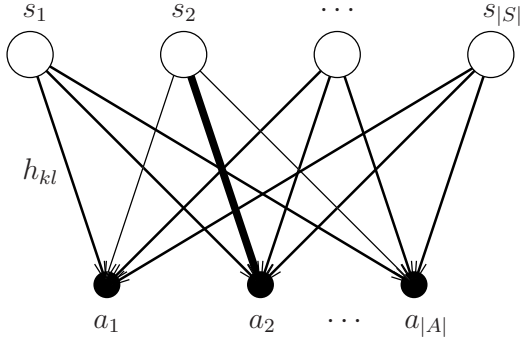}
\caption{\label{fig3}
Transition strengths ${h}_{kl}$ for an invasion game-like task with states
(which are here synonymous to percept-clips) $s_k$ and action clips $a_l$. The
strengthening of a rewarded transition (here ${h}_{22}$) is in an update
(\ref{GMcomp}) based on the gradient of $p(a_{j=2}|s_{i=2})$ accompanied by a
weakening of the respective transitions to the remaining actions
(here ${h}_{21}$ and ${h}_{23}$), which is absent in PS, cf. (\ref{PScomp}).
}
\end{figure}

Our method (\ref{GM1})-(\ref{GM2}) replaces a value function with a conditional
probability (\ref{grada}), whereas the edge strengths in PS remain value
function-like quantities. While tabular RL can be recovered from gradient-ascent
RL, one hence cannot expect to recover the basic PS update rule
(\ref{PS1})-(\ref{PS3}) as a special case of our scheme, despite replacements
analogous to (\ref{update1})-(\ref{update2}). To understand the difference, we
restrict attention to an invasion game - like case as shown in
Fig.~\ref{fig3} as the simplest example, cf. also Sec.~\ref{sec4} for
details.
Since here, each episode lasts one cycle, we disable both the eligibility
trace/glow mechanism by setting $\eta$ $\!=$ $\!1$ in (\ref{GM1})-(\ref{GM2})
and (\ref{PS1})-(\ref{PS3}). As shown in Fig.~\ref{fig3}, we are given a set
of states $\{s_1,s_2,\ldots,s_{|S|}\}$, each of which allows one out of a set of
actions $\{a_1,a_2,\ldots,a_{|A|}\}$. Consider a cycle, where from state $s_i$,
an action $a_j$ is selected. If the transition probabilities
\begin{equation}
  p_{ij}=p(a_j|s_i)=\frac{\Pi({h}_{ij})}{c_i},\quad{c}_i=\sum_j\Pi({h}_{ij}),
\end{equation}
are given by some policy function $\Pi$, then the components of the r.h.s. of
(\ref{GM1}) read
\begin{equation}
\label{eq23}
  e_{kl|ij}=\frac{\partial{p}_{ij}}{\partial{h}_{kl}}
  =\frac{\delta_{ik}\Pi^\prime({h}_{kl})}{c_i^2}
  \bigl[\delta_{jl}c_i-\Pi({h}_{ij})\bigr],
\end{equation}
with which the components of the update (\ref{GM2}) become
\begin{equation}
  {h}_{kl}\leftarrow
  (1-\kappa){h}_{kl}+\alpha{r}e_{kl|ij}+\kappa{h}_{kl}^{\mathrm{eq}},
\end{equation}
where we have renamed $\bm{h}_\infty$ as $\bm{h}^{\mathrm{eq}}$. An observer
ignorant of the transitions $i$ $\!\to$ $\!j$ and the corresponding
probabilities $p_{ij}$ notices no change,
\begin{equation}
  \sum_{ij}e_{kl|ij}=0.
\end{equation}
In the special case $\Pi({h}_{kl})$ $\!=$ $\!{h}_{kl}$, we can simplify
(\ref{eq23}) to
\begin{equation}
\label{GMcomp}
  e_{kl|ij}=\frac{1}{c_i}\left(\delta_{ik}\delta_{jl}-
  \frac{{h}_{ij}}{c_i}\delta_{ik}\right).
\end{equation}
From (\ref{GMcomp}) we see that in the gradient method, the strengthening of the
${h}_{ij}$-edge is accompanied with a weakening of those edges ${h}_{kl}$
connecting the respective state $k$ $\!=$ $\!i$ with different actions 
$l$ $\!\neq$ $\!j$. As a consequence, the ${h}_{kl}$ may become negative, even
if $\Pi({h}_{kl})$ $\!=$ $\!{h}_{kl}$ and the rewards are non-negative. This
weakening is absent in basic PS (\ref{PS2}), where the corresponding update is
independent of the policy function $\Pi$ and given by
\begin{eqnarray}
  {h}_{kl}&\leftarrow&
  {h}_{kl}-\gamma({h}_{kl}-{h}_{kl}^{\mathrm{eq}})
  +\lambda\delta_{ik}\delta_{jl}
  \nonumber\\
  &=&(1-\gamma){h}_{kl}+\lambda\delta_{ik}\delta_{jl}
  +\gamma{h}_{kl}^{\mathrm{eq}}.
\label{PScomp}
\end{eqnarray}
Hence, ${h}_{kl}$ $\!\ge$ $\!0$ as long as the rewards are non-negative. In any
case, choice of a non-negative policy function $\Pi({h}_{kl})$ renders the
methods independent of a need of positive edge strengths. [Note that a similar
problem occurs if the parameters in a memory consisting of alternating layers
such as $\hat{U}$ $\!=$ $\!\cdots\mathrm{e}^{-\mathrm{i}{t}_3\hat{H}^{(1)}}
\mathrm{e}^{-\mathrm{i}{t}_2\hat{H}^{(2)}}\mathrm{e}^{-\mathrm{i}{t}_1
\hat{H}^{(1)}}$, cf. App.~\ref{app:fl}, refer to non-negative physical
quantities $t_k$. In \cite{Cla15}, this has been solved by using an exponential
function such as ${t}_k$ $\!=$ $\!\mathrm{e}^{{h}_k}$ for parametrisation in
terms of the controls $h_k$. In this work, we identify the $h_k$ directly with
the $t_k$ for simplicity, which, if the $h_k$ are to be interpreted as
non-negative quantities, doubles the set of physically applied Hamiltonians from
$\{\hat{H}^{(1,2)}\}$ to $\{\pm\hat{H}^{(1,2)}\}$.]
\subsection{Discussion}
\begin{figure}[ht]
\includegraphics[width=7cm]{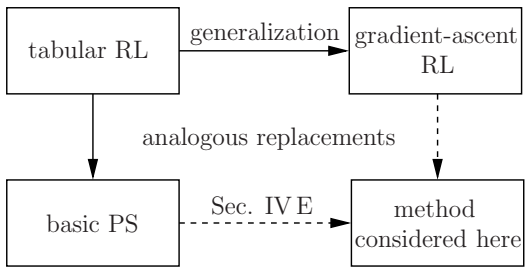}
\caption{\label{fig4}
Tabular RL (\ref{TRL1})-(\ref{TRL3}) is a special case of gradient-ascent RL
(\ref{GRL1})-(\ref{GRL2}). Replacing updates based on values of subsequent
states with updates based on a physical damping term yields basic
PS (\ref{PS1})-(\ref{PS3}) and the method presented here
(\ref{GM1})-(\ref{GM2}), which however uses a conditional probability
(\ref{grada}) instead of a value function, hence the basic PS update rule cannot
be recovered from our approach, as explained in Sec.~\ref{sec:VE}.
}
\end{figure}

The relations between the different methods are summarised in
Fig.~\ref{fig4}.
If one considers the ECM as the core element of PS rather than a specific update
rule, one could alternatively adopt, e.g., the tabular SARSA($\lambda$)-update
rule. The picture of a random walk in clip space does not contradict the general
framework of RL-problems. One may understand the clips as the agent's states
(which must be distinguished from the percepts). The same holds for the
gradient-ascent generalization, which, in physical terms, could be considered as
``continuous variable RL''. On the one hand, we could equally well apply, e.g.,
the gradient-ascent SARSA($\lambda$)-update instead of our rule. On the other
hand, before trying to create algorithmic extensions such as those mentioned at
the beginning of this section for tabular RL and PS, one should first
investigate whether and how such extensions are accomplished in any existing
gradient-ascent RL variants.
\section{
\label{sec4}
Examples
}
\subsection{
\label{sec:ig}
Invasion game}
In what follows, we consider a simple invasion game as treated in \cite{Bri12}.
An attacker randomly chooses one out of two possible symbols
$\{\Leftarrow,\Rightarrow\}$ which signals the direction in which it intends to
move. The chosen symbol may represent, e.g., a head turn and is visible to the
defender, whose task is to learn to move in the same direction, which is
required to block the attacker. We approach this learning task as an external
loop in Fig.~\ref{fig1} with a closed system (i.e., bath-less) memory [cases
(a) and (c) in Fig.~\ref{fig2}], described within a 4-dimensional Hilbert
space. The control parameters are updated according to (\ref{ur3}) in the
absence of relaxation ($\kappa$ $\!=$ $\!0$) and gradient glow
($\eta$ $\!=$ $\!1$). The update is done with an analytic $\bm{\nabla}\hat{U}$
as described in App.~\ref{app:fl}, where the memory consists of alternating
layers, $\hat{U}$ $\!=$ $\!\cdots\mathrm{e}^{-\mathrm{i}{h}_3\hat{H}^{(1)}}
\mathrm{e}^{-\mathrm{i}{h}_2\hat{H}^{(2)}}\mathrm{e}^{-\mathrm{i}{h}_1
\hat{H}^{(1)}}$, with a given number of controls $h_1,\ldots,h_n$. At the
beginning of the first cycle, the memory is initialised as identity. For the two
Hamiltonians $\hat{H}^{(1)}$ and $\hat{H}^{(2)}$, we distinguish
(I) a general case, where $\hat{H}^{(1)}$ and $\hat{H}^{(2)}$ are two given
(randomly generated) 4-rowed Hamiltonians acting on the total Hilbert space and
(II) a more specialised case, in which they have the form
\begin{eqnarray}
\label{HA}
  \hat{H}^{(1)}&=&\hat{H}^{(1)}_{\mathrm{S}}\otimes\hat{I}_{\mathrm{A}}
  +\hat{I}_{\mathrm{S}}\otimes\hat{H}^{(1)}_{\mathrm{A}},
  \\
\label{HB}
  \hat{H}^{(2)}&=&\hat{H}^{(2)}_{\mathrm{S}}\otimes\hat{H}^{(2)}_{\mathrm{A}},
\end{eqnarray}
where $\hat{H}^{(1)}_{\mathrm{S}}$, $\hat{H}^{(1)}_{\mathrm{A}}$,
$\hat{H}^{(2)}_{\mathrm{S}}$, $\hat{H}^{(2)}_{\mathrm{A}}$ are four given
(randomly generated) 2-rowed Hamiltonians acting on the percept (S) and action
(A) subsystems, respectively, with $\hat{I}$ denoting the identity. The latter
case (II) refers to a physical implementation of Fig.~\ref{fig2}(c) as a
bath-mediated interaction of the S and A subsystems that is obtained from the
setup Fig.~\ref{fig2}(d) by eliminating the bath \cite{Cla15}. It has been
included here to demonstrate that this special structure as considered in
\cite{Cla15} may be applied in the present context, but this is not mandatory.
While the Hamiltonians have been chosen in both cases (I) and (II) at random to
avoid shifting focus towards a specific physical realization, in an experimental
setup, the respective laboratory Hamiltonians will take their place (assuming
that they generate universal gates in the sense of \cite{lloyd2}, which is
almost surely the case for a random choice).
\subsubsection{
\label{sec:ig22}
2 percepts $\to$ 2 actions}
We start with a basic version of the game with 2 possible percepts (the two
symbols shown by the attacker) and 2 possible actions (the two moves of the
defender). For each percept, there is hence exactly one correct action, which is
to be identified. The memory applied is shown in Fig.~\ref{fig2}(c), and the
different input states are
\begin{eqnarray}
\label{is}
  \hat{\varrho}&=&\hat{\varrho}_{\mathrm{S}}\otimes\hat{\varrho}_{\mathrm{A}},
  \quad
  \hat{\varrho}_{\mathrm{S}}=|{s}\rangle\langle{s}|,
  \\
\label{pcoh}
  \hat{\varrho}_{\mathrm{A}}&=&p_{\mathrm{coh}}|\varphi\rangle\langle\varphi|
  +(1-p_{\mathrm{coh}})\frac{1}{d_{\mathrm{A}}}\hat{I}_{\mathrm{A}},\quad
  \\
\label{ia}
  |\varphi\rangle&=&\frac{1}{\sqrt{d_{\mathrm{A}}}}\sum_a|{a}\rangle,
\end{eqnarray}
where
$d_{\mathrm{A}}$ $\!=$ $\!\mathrm{dim}\mathcal{H}_{\mathrm{A}}$ $\!=$ $\!2$ is
given by the number of actions. $|s\rangle$ and $|a\rangle$ can both be one of
the two orthonormal states $|0\rangle$ or $|1\rangle$ of the S and A subsystem,
respectively. The POVM consists of the elements
\begin{equation}
\label{pom}
  \hat{\Pi}(a)=\hat{I}_{\mathrm{S}}\otimes|{a}\rangle_{\mathrm{A}}\langle{a}|.
\end{equation}
Choosing the correct (wrong) action [i.e. $a$ $\!=$ $\!s$ ($a$ $\!\neq$ $\!s$)
in (\ref{pom}) and (\ref{is})] returns a reward of
$r$ $\!=$ $\!+1$ ($r$ $\!=$ $\!-1$).

\begin{figure}[ht]
\includegraphics[width=4.2cm]{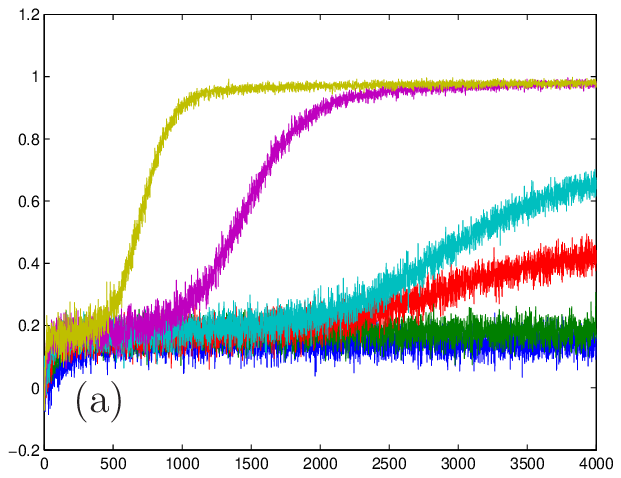}
\includegraphics[width=4.2cm]{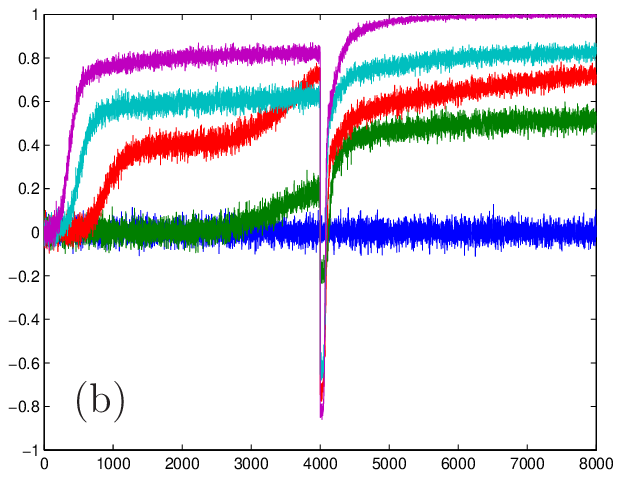}
\includegraphics[width=4.2cm]{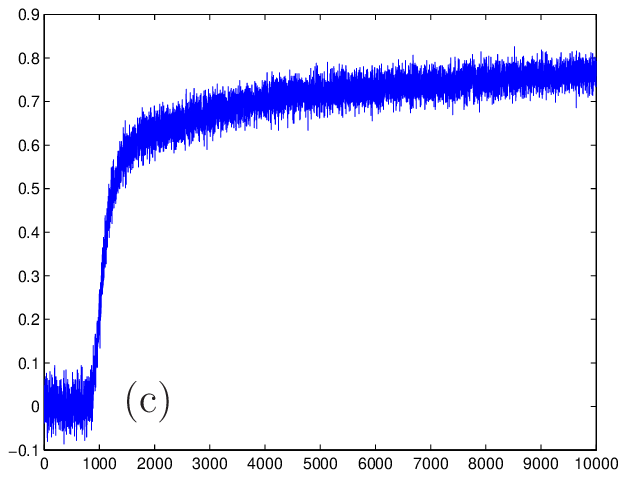}
\caption{\label{fig5}
Average reward $\overline{r}$ as a function of the number of cycles for an
invasion game (2 symbols, 2 moves), learning rate $\alpha$ $\!=$ $\!10^{-3}$,
with a reward of +1 (-1) for a correct (false) move, averaged over
$10^3$ agents. The input states and POVM are given by Eqs.~(\ref{is}) and
(\ref{pom}), respectively.
(a) The 6 graphs from bottom to top correspond to 1, 2, 3, 4, 8, and 16
controls, respectively, with $p_{\mathrm{coh}}$ $\!=$ $\!1$ in (\ref{pcoh}).
(b) 16 controls, where after $4\cdot10^3$ cycles the meaning
of the symbols is reversed. The 5 graphs from bottom to top correspond in
Eq.~(\ref{pcoh}) to $p_{\mathrm{coh}}$= 0, 0.25, 0.5, 0.75, and 1, respectively.
(c) applies 32 controls and $p_{\mathrm{coh}}$ $\!=$ $\!1$, but refers to case
(II) described by (\ref{HA}) and (\ref{HB}), whereas Figs.~\ref{fig5}(a,b)
refer to case (I).
}
\end{figure}

Fig.~\ref{fig5} shows the average reward $\overline{r}$ received at each
cycle, where the averaging is performed over an ensemble of $10^3$ independent
agents.
$(\overline{r}+1)/2$ is hence an estimate of the defender's probability to
block an attack. Referring to pure states in (\ref{pcoh}), Fig.~\ref{fig5}(a)
shows the increase of learning speed with the number of controls. Significant
learning progress begins only after some initial period of stagnation. From the
viewpoint of control theory, the identity, in which the memory is initialised,
may lie on a ``near-flat ground'' (valley), which must first be left before
progress can be made \cite{Goe15}. Asymptotically, perfect blocking can be
achieved once the memory becomes controllable, i.e., if the number of controls
equals (or exceeds) the number of group generators.
Fig.~\ref{fig5}(b) demonstrates the need of a pure input state
$\hat{\varrho}_{\mathrm{A}}$ in (\ref{pcoh}) of the action subsystem A rather
than an incoherent mixture. After the agent in Fig.~\ref{fig5}(b) had some time
to adapt to the attacker, the meaning of the symbols is suddenly interchanged,
and the agent must now learn to move in the opposite direction. This relearning
differs from the preceding learning period in the absence of the mentioned
initial stagnation phase, which supports the above hypothesis of the proposed
valley, the agent has left during the initial learning. This plot is motivated
by Fig. 5 in \cite{Bri12} describing classical PS. Although the different
behaviour in the classical case suggests that this is an effect specific to
quantum control, the phenomenon, that a dynamically changing environment can
facilitate learning in later stages appears to be more general \cite{z3}. While
Figs.~\ref{fig5}(a,b) refer to case (I) described before (\ref{HA}),
Fig.~\ref{fig5}(c) refers to the restricted case (II), which appears to impede
learning. In the simulations of the following Sec.~\ref{sec:ig44}, which all
refer to case (II), this is resolved by applying a 10 times larger negative
reward for each wrong action. This demonstrates the flexibility in approaching
RL problems offered by the freedom to allocate rewards in a suitable way.
\subsubsection{
\label{sec:ig44}
4 percepts $\to$  4 or 2 actions}
\begin{figure}[ht]
\hspace{0.7cm}\includegraphics[width=6.7cm]{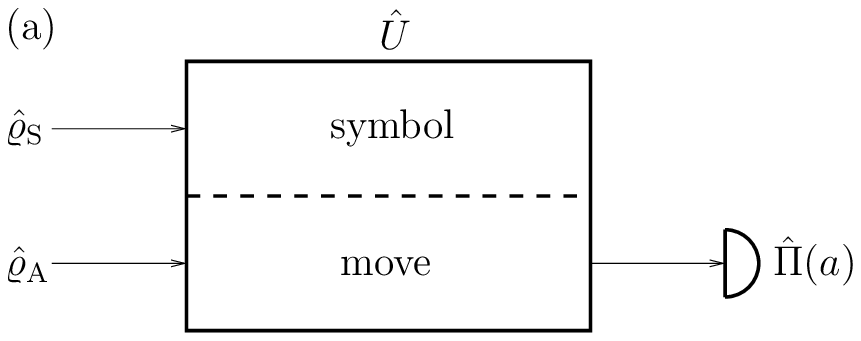}\\
\vspace{0.5cm}
\includegraphics[width=6cm]{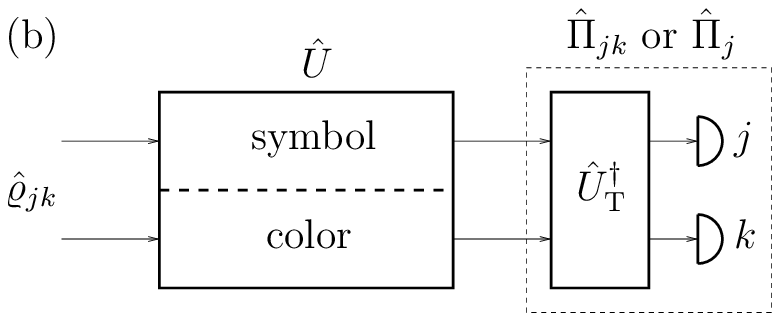}\\
\vspace{0.5cm}
\hspace{0.7cm}\includegraphics[width=6.7cm]{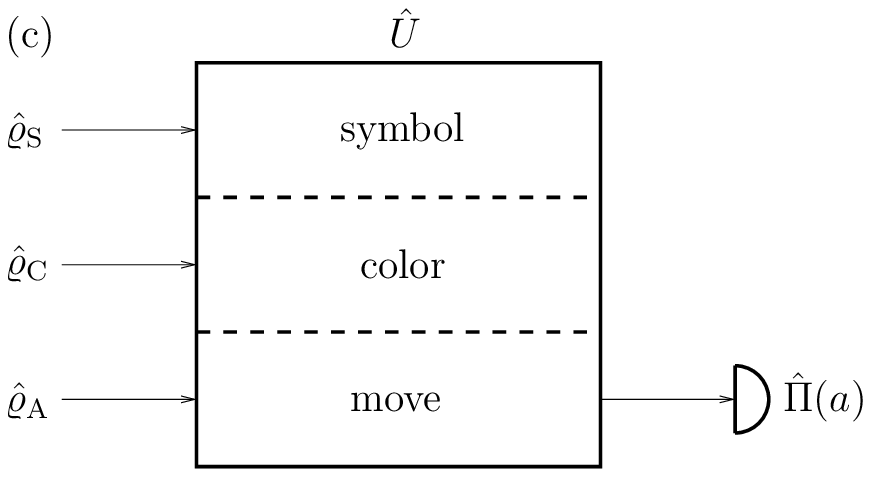}
\caption{\label{fig6}
Setups for the invasion game as investigated numerically in Sec.~\ref{sec:ig}.
Setup (a) involves 2 percepts (symbols) and 2 actions (moves) as discussed in
Sec.~\ref{sec:ig22} (Fig.~\ref{fig5}). Setup (b) involves 4 percepts consisting
of 2 two-colored symbols and 2 measurements yielding 4 different outcomes
described by $(j,k=0,1)$ as discussed in Sec.~\ref{sec:ig44} and determining
either 4 [Fig.~\ref{fig7}(a) and Fig.~\ref{fig8}] or -- if outcome $k$ is
ignored -- 2 [Fig.~\ref{fig7}(b)] possible actions (moves). While setup (a)
refers to Fig.~\ref{fig2}(c), setup (b) must refer to Fig.~\ref{fig2}(a), if
we want to keep the same Hilbert space dimension of 4 for both setups, which
allows better comparison of the effect of the number of controls on the learning
curve. Setup (c) involves a continuum of percepts consisting of 2
arbitrary-colored symbols and 2 actions (moves) as discussed in
Sec.~\ref{sec:nec} (Fig.~\ref{fig9}). In setup (c), separate subsystems are
used for all 3 categories, hence it refers to Fig.~\ref{fig2}(c), and the
Hilbert space dimension becomes 8.
}
\end{figure}

We now consider a version with 4 percepts, referring to an attacker presenting
each of its two symbols in two colors at random. Since we want to keep the
Hilbert space dimension unchanged (rather than doubling it by adding the color
category) for better comparison of the effect of the number of controls on the
learning curve, we must apply a memory as shown in Fig.~\ref{fig2}(a).
The 4 percepts are encoded as tensor products of orthonormal projectors
\begin{equation}
\label{is4}
  \hat{\varrho}_{jk}=|{j}\rangle\langle{j}|\otimes|{k}\rangle\langle{k}|,
\end{equation}
where $j$ $\!=$ $\!0,1$ ($k$ $\!=$ $\!0,1$) refers to the symbol (color).
The POVM operators are the 4 projectors
\begin{equation}
\label{pom4}
  \hat{\Pi}_{jk}=\hat{U}_{\mathrm{T}}\hat{\varrho}_{jk}
  \hat{U}_{\mathrm{T}}^\dagger,
\end{equation}
where $\hat{U}_{\mathrm{T}}$ is a given (randomly generated) 4-rowed target
unitary acting on the total system. The memory in Fig.~\ref{fig2}(a) is hence
still composed of two subsystems referring to the two percept categories
`symbol' and `color', but both subsystem's initial state depends on the
respective percept, and both are measured afterwards. The differences between
the setup discussed in the previous Sec.~\ref{sec:ig22} and the two setups
discussed in the present Sec.~\ref{sec:ig44} are summarised in Fig.~\ref{fig6}.

\begin{figure}[ht]
\includegraphics[width=4.2cm]{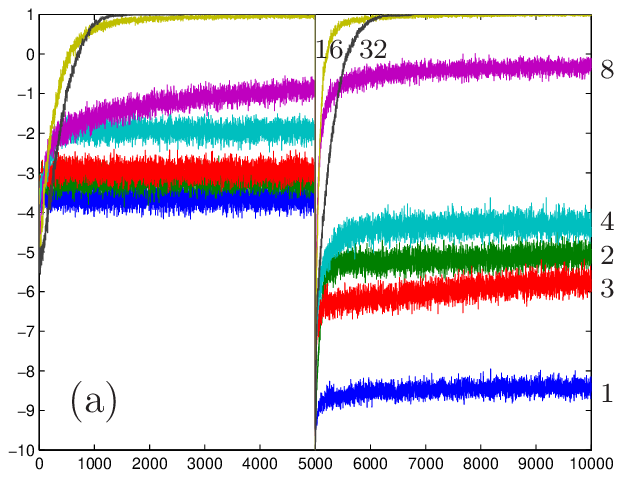}
\includegraphics[width=4.2cm]{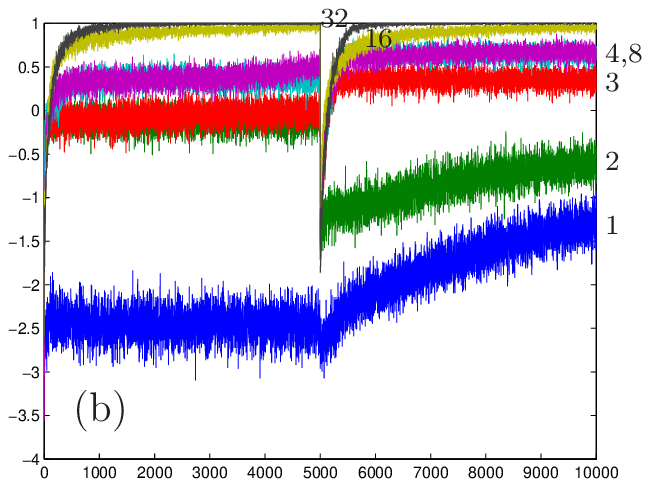}
\caption{\label{fig7}
Average reward $\overline{r}$ as a function of the number of cycles for an
invasion game (2 symbols in 2 colors), learning rate $\alpha$ $\!=$ $\!10^{-2}$,
with a reward of +1 (-10) for a correct (false) move, averaged over
$10^3$ agents. The input states and POVM are given by Eqs.~(\ref{is4}) and
(\ref{pom4}), respectively. The graphs correspond to 1, 2, 3, 4, 8, 16, and 32
controls, as marked on the right.
(a) Out of 4 possible moves, the defender must learn the correct one for each
symbol and color. After $5\cdot10^3$ cycles, the meanings of the symbols as well
as the colors are reversed.
(b) Out of 2 possible moves, the defender must learn the correct one for each
symbol, whereas the color is irrelevant. For the first $5\cdot10^3$ cycles, only
symbols in a single color are presented, whereas for the remaining cycles, they
are shown randomly in both colors.
}
\end{figure}

Fig.~\ref{fig7} shows the average reward $\overline{r}$ received at each
cycle, where the averaging is performed over an ensemble of $10^3$ independent
agents, analogous to Fig.~\ref{fig5}.
Note that in this Sec.~\ref{sec:ig44}, all figures refer to case (II) described
by (\ref{HA}) and (\ref{HB}), where S and A now denote symbol and color,
respectively. To account for this [cf. the comments on Fig.~\ref{fig5}(c)
above], a reward of $r$ $\!=$ $\!-10$ (instead of -1) is now given for a wrong
action. The estimate of the defender's probability to block an attack is hence
now $(\overline{r}+10)/11$.

In Fig.~\ref{fig7}(a), the defender can choose between 4 moves, where for each
percept, there is exactly one correct action [i.e., detecting $\hat{\Pi}_{jk}$
($\hat{\Pi}_{j^\prime{k}^\prime}$ $\!\neq$ $\!\hat{\Pi}_{jk}$) for
$\hat{\varrho}_{jk}$ in (\ref{pom4}) and (\ref{is4}) returns a reward of
$r$ $\!=$ $\!+1$ ($r$ $\!=$ $\!-10$)]. After $5\cdot10^3$
cycles, symbol $j$ and color $k$ are read as symbol $1-j$ and color $1-k$,
respectively, similar to the manipulations in Fig. 5 in \cite{Bri12}.
In Fig.~\ref{fig7}(b), the defender can choose between 2 moves, where for each
symbol (relevant category), there is exactly one correct action, irrespective
of its color (irrelevant category) [i.e., detecting
$\hat{\Pi}_{j}$ $\!=$ $\!\sum_{k=0}^{1}\hat{\Pi}_{jk}$
($\hat{\Pi}_{j^\prime}$ $\!\neq$ $\!\hat{\Pi}_{j}$) for $\hat{\varrho}_{jk}$ in
(\ref{pom4}) and (\ref{is4}) returns a reward of
$r$ $\!=$ $\!+1$ ($r$ $\!=$ $\!-10$)]. The second color is added only after 
$5\cdot10^3$ cycles, analogous to Fig. 6 in \cite{Bri12}.
[Note that the mentioned initial stagnation phase in Fig.~\ref{fig5} is not
visible in Fig.~\ref{fig7}, which is attributed to the choice of parameters
(rewards), accelerating the initial learning.]

Figs.~\ref{fig5}(b) and \ref{fig7} are all motivated by Figs. 5 and 6 in
\cite{Bri12} and confirm that the agent's adaptation to changing environments
is recovered in our quantum control context. In addition, Figs.~\ref{fig5}(a)
and \ref{fig7} show the behaviour of an underactuated memory, where the number
of controls is insufficient for its full controllability. Since a $U(n)$-matrix
is determined by $n^2$ real parameters, and a global phase can be disregarded
(so that we can restrict to SU($n$)-matrices), $n^2$ $\!-$ $\!1$ controls are
sufficient, i.e., 15 for our invasion game, as mentioned above.

\begin{figure}[ht]
\includegraphics[width=4.2cm]{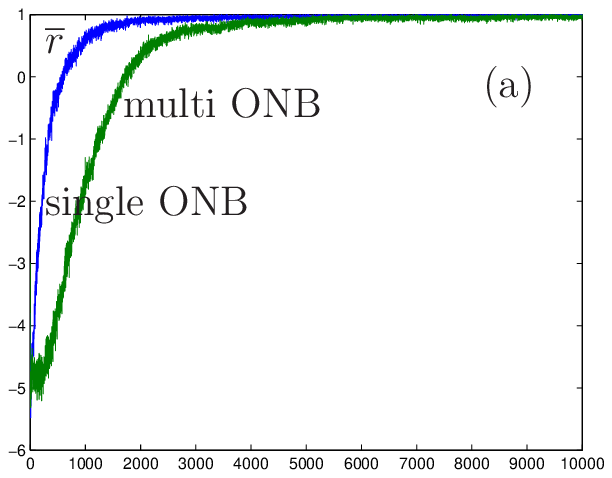}
\includegraphics[width=4.2cm]{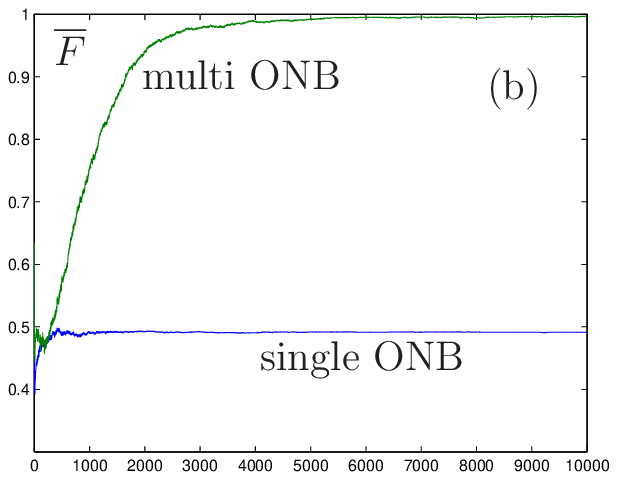}
\includegraphics[width=4.2cm]{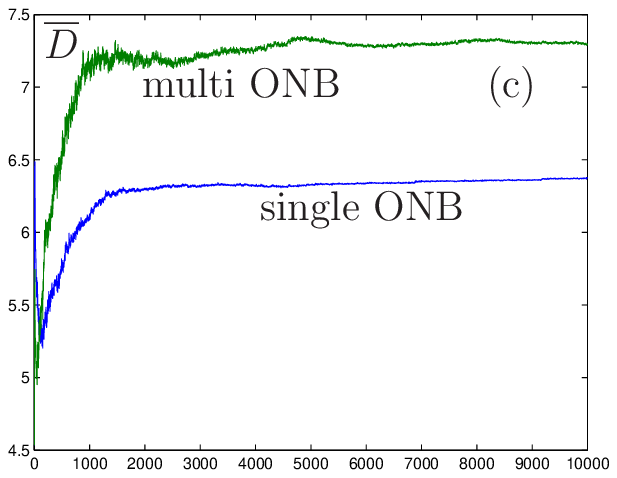}
\caption{\label{fig8}
(a) Reward $r$, (b) fidelity $F$ defined in (\ref{F}), and (c) squared distance
$D$ defined in (\ref{D}), where the overline denotes the ensemble average over
$10^3$ agents for the setup as in Fig.~\ref{fig7}(a) (i.e., 4 percepts and 4
actions) with 16 controls but without the reversal of meaning. The initial
memory states are drawn from either a single or multiple orthonormal bases.
}
\end{figure}

In (\ref{pom4}), the measurements are made in a basis rotated by a randomly
given unitary $\hat{U}_{\mathrm{T}}$, which serves two purposes. On the one
hand, it is required to ensure that the agent starts at the beginning of the
first cycle with a policy that does not give exclusive preference to certain
actions that follow from symmetries of the (identity-) initialised memory. This
is a flaw of Fig.~\ref{fig2}(a) and can be overcome by using
Fig.~\ref{fig2}(c) instead (cf. a more detailed discussion in the grid world
example below). On the other hand, $\hat{U}_{\mathrm{T}}$ serves as a given
target in our discussion Sec.~\ref{sec3}, where we consider the agent learning
as a navigation of its memory $\hat{U}$, cf. also Fig.~\ref{fig6}(b).
Fig.~\ref{fig8} compares the case, where the agent is always fed with percept
states drawn from one single ONB defined via (\ref{is4}) with the case, where
the percept states are drawn randomly, i.e., taking
$\hat{U}_{\mathrm{R}}\hat{\varrho}_{jk}\hat{U}_{\mathrm{R}}^\dagger$ with a
random unitary $\hat{U}_{\mathrm{R}}$ as explained in Sec.~\ref{sec3} instead of
(\ref{is4}). Note that in Fig.~\ref{fig8}, we generate a new random
$\hat{U}_{\mathrm{R}}$ at each cycle, although a fixed set of dim$\mathcal{H}$
(4 in our case) such $\hat{U}_{\mathrm{R}}$ is sufficient as mentioned in
Sec.~\ref{sec3}.
Fidelity $F$ and squared distance $D$ are defined in (\ref{F}) and (\ref{D}),
where $\hat{U}$ represents the agent memory and $\hat{U}_{\mathrm{T}}$ the
target unitary. Each cycle's update constitutes a single navigation step in the
unitary group [U(4) in our example]. If, for a single ONB, after a number of
cycles, the average reward has approached unity, $\hat{U}$ has reached a close
neighbourhood of any unitary of the form
$\hat{U}_2^\prime\hat{U}_{\mathrm{T}}\hat{U}_1$,
where $\hat{U}_2^\prime$ $\!=$
$\!\hat{U}_{\mathrm{T}}\hat{U}_2\hat{U}_{\mathrm{T}}^\dagger$
with $\hat{U}_1$ and $\hat{U}_2$ being undetermined 4-rowed unitary matrices
diagonal in the common eigenbasis of the $\hat{\varrho}_{jk}$
(i.e., the ``computational basis''). Fig.~\ref{fig8}(a) shows that for a
solution of the invasion game, a fixed ONB is sufficient. Drawing the percept
states randomly, so that the set of all percept states is linearly dependent,
does not affect the agent's ability to achieve perfect blocking efficiency, but
slows down the learning process. The single ONB case allows for a larger set of
$\hat{U}$ $\!=$ $\!\hat{U}_2^\prime\hat{U}_{\mathrm{T}}\hat{U}_1$ with respect
to $\hat{U}_{\mathrm{T}}$, as becomes evident in Fig.~\ref{fig8}(b), so that
navigation of $\hat{U}$ from the identity to a member of this set takes less
time (as measured in cycles). The only freedom left in the case of multiple ONBs
is a global phase of $\hat{U}$, which remains undefined: navigation of $\hat{U}$
towards $\hat{U}_{\mathrm{T}}$ with respect to the squared Euclidean distance
$D$ is not required for the learning tasks discussed, as evidenced by
Fig.~\ref{fig8}(c).
\subsubsection{
\label{sec:nec}
Neverending-color scenario}
In Sec.~\ref{sec:ig44} we considered the case, where the symbols are presented
in two different colors, as depicted in Fig.~\ref{fig6}(b). The original
motivation for introducing colors as an additional percept category was to
demonstrate the agent's ability to learn that they are irrelevant \cite{Bri12}.
In contrast, \cite{Mel15} present a ``neverending-color scenario'', where at
each cycle, the respective symbol is presented in a new color. It is shown that
while the basic PS-agent is in this case unable to learn at all, it becomes able
to \emph{generalize} (abstract) from the colors, if it is enhanced by a
wildcard mechanism. The latter consists in adding an additional
(wildcard ``$\#$'') value to each percept category, and inserting between the
input layer of percept clips and the output layer of action clips hidden layers
of wildcard percept clips, in which some of the percept categories attain the
wildcard value. The creation of these wildcard clips follows predefined
deterministic rules, and the transitions from percept to action clips take then
place via the hidden layers. (The notion of layers in the general ECM clip
network Fig. 2 in \cite{Bri12} follows from restricting to clips of length
$L$ $\!=$ $\!1$).

Since the use of wildcard clips is an integrated mechanism within PS (inspired
by learning classifier systems), the question is raised  how similar ideas could
be implemented in our context. For a memory Fig.~\ref{fig2}(c), we could, e.g.,
attribute one of the levels (such as the respective ground state) of the quantum
system of each percept category $\mathrm{S}_i$ to the wildcard-level
$|\#\rangle_i$, so that the percept space
$\mathcal{H}_{\mathrm{S}}$ $\!=$ $\!\otimes\mathcal{H}_{\mathrm{S}_i}$ is
enlarged to
$\mathcal{H}_{\mathrm{S}}$ $\!=$ $\!\otimes(\mathcal{H}_{\mathrm{S}_i}$
$\!\oplus$ $\!\mathcal{H}_{\#_i})$, where the $\mathcal{H}_{\#_i}$ are
one-dimensional.

\begin{figure}[ht]
\includegraphics[width=4.2cm]{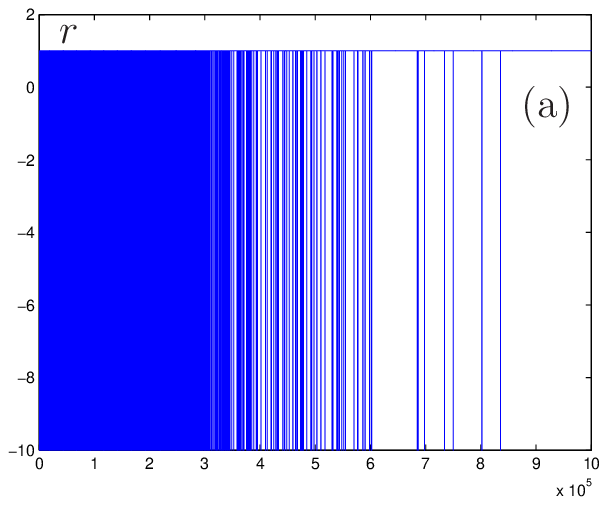}
\includegraphics[width=4.2cm]{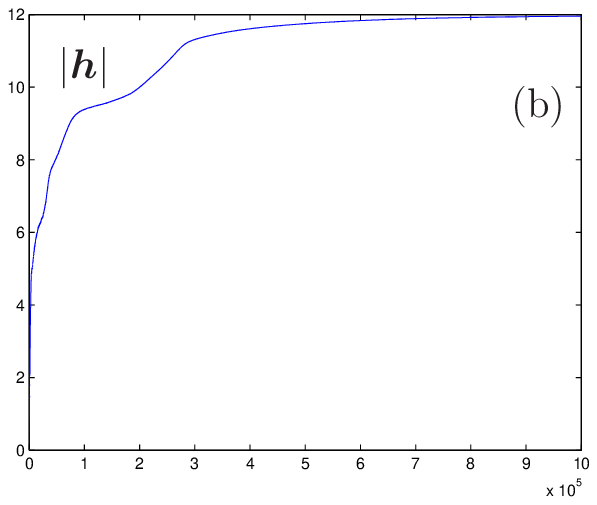}
\caption{\label{fig9}
(a) Reward $r$ and (b) length $|\bm{h}|$ of the control vector as a function of
the number of cycles for a single agent Fig.~\ref{fig6}(c) playing an invasion
game with 2 symbols, presented in a continuum of neverending colors, and 2
moves. A reward of +1 (-10) is given for each correct (false) move. The agent
applies 64 controls with a learning rate of $\alpha$ $\!=$ $\!10^{-2}$ in an
alternating layer scheme App.~\ref{app:fl} defined by two
(Schmidt-orthonormalized) 8-rowed random Hamiltonians.
}
\end{figure}

Instead of this, let us simply make use of the \emph{built-in} generalization
capacity of a quantum agent resulting from its coding of percepts as quantum
states, which is much in the sense of Sec. 8 in \cite{bookSuttonBarto}, where
the percepts can be arbitrarily real-valued rather than being drawn from a
countable or finite value set. Consider the setup shown in Fig.~\ref{fig6}(c),
whose percept system includes a symbol and a color category and refers to a
memory structure Fig.~\ref{fig2}(c). To allow for infinite colors, we could
apply a color quantum system with infinite levels $\mathcal{H}_\infty$  (such as
an oscillator-type system), which is initialized at each cycle in a new state
drawn from a fixed ONB (such as a higher number state for an oscillator-type
system). While such a scheme becomes more challenging to control, because the
control vector $\bm{h}$ has an infinite number of components [we may replace it
with a continuous control function $h(t)$], it still ignores the fact that
colors (as most percept categories in general) are not countable. With this in
mind, we can take the notion of colors literally and, to put it simply, code
them in some appropriate color space such as RGB, where three parameters
correspond to the red-, green-, and blue-signals of the agent's eye sensors.
This suggests to encode a color as a mixed state of a two-level system, which is
also given by three real-valued parameters (determining its location in the
Bloch ball). The generalization from two colors to all RGB-colors then
corresponds to the generalization from a classical to a quantum bit. In our
setup, it is hence sufficient to apply a two-level system for the color category
and initialize it at each cycle in a randomly chosen \emph{mixed} state
$\hat{\varrho}_{\mathrm{C}}$ (for neverending colors) rather than a (pure) state
randomly drawn from a single ONB (for two colors), whereas no changes are
required on the agent's memory configuration itself. Fig.~\ref{fig9}
demonstrates the learning process.
Similar to Fig.~\ref{fig8}, random initialization slows down the learning
process, so that we restrict to a single agent in Fig.~\ref{fig9}, rather than
an ensemble average. As illustrated in Fig.~\ref{fig9}(a), the agent's response
becomes near-deterministic after about $10^6$ cycles, irrespective of the color.
Fig.~\ref{fig9}(b) illustrates in the example of the Euclidean length of the
control vector $|\bm{h}|$ $\!=$ $\!\sqrt{\bm{h}^\mathrm{T}\cdot\bm{h}}$, that
the navigation, which starts at $\bm{h}_0$ $\!=$ $\!\bm{0}$, eventually comes to
rest. While the random $\hat{\varrho}_{\mathrm{C}}$ are drawn such that a
positive probability is attributed to every volume element in the Bloch ball, we
did not care about drawing them with a uniform probability density, since
mapping of an RGB-space of color (as a perceptual property) to the Bloch ball
is not uniquely defined.

The ability to learn to distinguish between relevant
$\mathrm{S}_{\mathrm{rel}}$ and an arbitrary number of irrelevant percept
categories $\mathrm{S}_{\mathrm{irr}}$ as discussed in \cite{Mel15} is of
particular relevance for a quantum agent, where the irrelevant percept
categories can be understood as adopting the role of a bath as shown in
Figs.~\ref{fig2}(b) and (d). Here, a formal solution consists in a decoupled
$\hat{U}_{\mathrm{SA}}$ $\!=$ $\!\hat{U}_{\mathrm{S_{rel}A}}$ $\!\otimes$
$\!\hat{U}_{\mathrm{S_{irr}}}$.
\subsection{
\label{sec:gw}
Grid world}
\begin{figure}[ht]
\includegraphics[width=6cm]{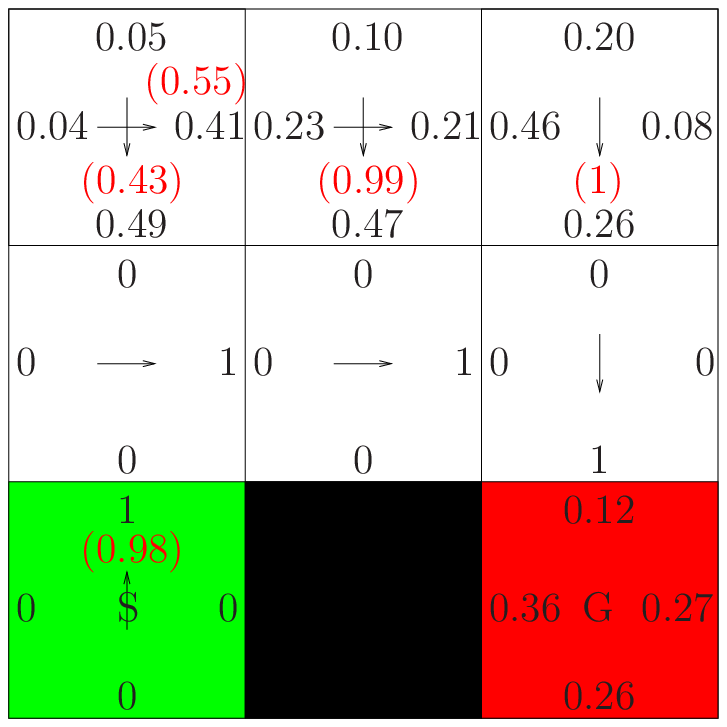}
\caption{\label{fig10}
$(3\times3)$-grid world with an obstacle (black cell). The arrows show the
optimal policy for the shortest path to G, and the numbers present the policy
numerically obtained with the agent Fig.~\ref{fig11}(d) after $10^{4}$
episodes. The red numbers in parentheses are the relevant different values
obtained after $10^{5}$ episodes, if the agent starts each episode from a random
cell, rather than always at S.
}
\end{figure}

In what follows, we consider an arrangement of 8 grid cells as shown in
Fig.~\ref{fig10}.
The agent's task is to find the shortest route to a goal cell G, where at each
step, only moves to an adjacent cell in four directions (left, right, up, down)
are allowed. If the agent hits a boundary of the grid or the black cell, which
is considered an obstacle, its location remains unchanged.
 
This external classical navigation task constitutes a learning problem, because
situations/percepts (present location) must be mapped to decisions/actions
(direction to go). The agent only perceives whether or not it has arrived at the
goal cell. It has no access to a ``bird's perspective'' which would allow
immediate exact location of the goal. It also has no access to a measure of goal
distance or fidelity (as in the case of the internal NO-based loop regarding its
own quantum memory in Fig.~\ref{fig1}), which prevents the use of external
gradient information that could be obtained by testing the nearest neighbourhood
of the present location. One can thus distinguish two objectives: (a) locating
the goal and (b) finding a shortest route to it. This task constitutes a RL type
problem, whose composite ``two-objective'' structure is approached by nesting
iterations. The individual action selections, i.e., choices of moves, correspond
to the cycles in Fig.~\ref{fig1}. Sequences of cycles form episodes, which are
terminated only once objective (a) has been solved. Objective (b) is solved by
sequences of episodes, which allow the agent to gradually solve objective (a)
more efficiently and find an optimal policy. In Fig.~\ref{fig10}, the policy
consists of a set of four probabilities for each cell, with which a
corresponding move should be made from there. The optimal policy corresponding
to the shortest route to G is indicated by the arrows in Fig.~\ref{fig10}.

This grid world extends the above decision game in two aspects:
(a) The optimal policy is in contrast to the decision game not deterministic,
as indicated by the double arrows in the upper left and middle cell in
Fig.~\ref{fig10}. (b) Depending on where the agent starts, more than a single
move is required to reach G in general, preventing the agent from obtaining an
immediate environmental feedback on the correctness of each individual move it
makes. This second aspect leads to the mentioned notion of episodes. In what
follows, we always place the agent at a fixed start cell S at the beginning of
each episode, which is sufficient for learning the shortest path from S to G.
While in the invasion game, episodes and cycles are synonyms, here, an episode
is longer than a single cycle, since at least four moves are required to reach G
from S.

When designing the agent's memory structure in the sense of Fig.~\ref{fig2},
we must take into account that the unitarity of the state transformation
$\hat{U}_{\mathrm{S}}$ in Fig.~\ref{fig2}(a) places restrictions on the
percept-encodings and the action-measurements, since $\hat{U}_{\mathrm{S}}$
maps an ONB into another one.
If we encode in Fig.~\ref{fig10} each cell location as a member of a given
ONB in an 8-dimensional system Hilbert space $\mathcal{H}_8$ and perform a naive
symmetric $\mathcal{H}_8$ $\!=$ $\!\mathcal{H}_2$ $\!\oplus$
$\!\mathcal{H}_2$ $\!\oplus$ $\!\mathcal{H}_2$ $\!\oplus$ $\!\mathcal{H}_2$
-measurement for action selection, where the four 2-dimensional subspaces
correspond to right, down, left and up moves, we cannot properly ascribe the
upper left and upper middle cells, because the right and downward pointing
actions are already ascribed to the remaining 4 white cells. One may either try
to construct a learning algorithm that exploits the fact that the two mentioned
cells are off the optimal path from S to G so that the agent quickly ceases to
visit them or construct a new POVM such as a
$\mathcal{H}_8$ $\!=$ $\!\mathcal{H}_1$ $\!\oplus$ $\!\mathcal{H}_3$ $\!\oplus$
$\!\mathcal{H}_3$ $\!\oplus$ $\!\mathcal{H}_1$-measurement, where two
3-dimensional subspaces correspond to right and down, and two 1-dimensional
subspaces correspond to left and up moves. These possibilities require insight
into the specifics of this problem and are not generalizable. In addition to
that, Fig.~\ref{fig2}(a) requires initialisation of the agent memory in a
random unitary to ensure it starts with a policy that does not give exclusive
preference to certain actions that follow from symmetries of the initial
$\hat{U}_{\mathrm{S}}$ (such as the identity
$\hat{U}_{\mathrm{S}}$ $\!=$ $\!\hat{I}_{\mathrm{S}}$). If we want to avoid
invoking a bath as in Fig.~\ref{fig2}(b), we hence must resort to
Fig.~\ref{fig2}(c), which here implies a factorisation
$\mathcal{H}$ $\!=$ $\!\mathcal{H}_{\mathrm{S}}$ $\!\otimes$
$\!\mathcal{H}_{\mathrm{A}}$ of $\mathcal{H}$ into an 8-dimensional
$\mathcal{H}_{\mathrm{S}}$ for encoding the grid cells and a 4-dimensional
$\mathcal{H}_{\mathrm{A}}$ for encoding the four actions. If we encode the cells
and actions as members of some ONB in S and A, then initialising the agent's
memory as identity, $\hat{U}_{\mathrm{SA}}$ $\!=$ $\!\hat{I}_{\mathrm{SA}}$,
and the initial action states as in (\ref{ia}) ensures that the agent starts at
the beginning of the first episode with a policy that assigns the same
probability to all possible actions.

\begin{figure}[ht]
\includegraphics[width=4.2cm]{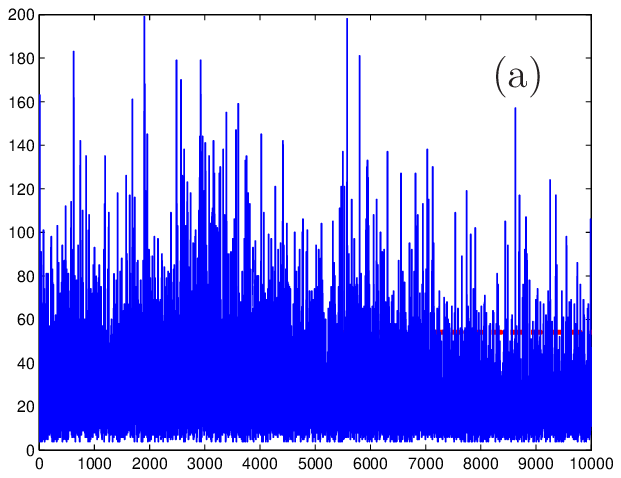}
\includegraphics[width=4.2cm]{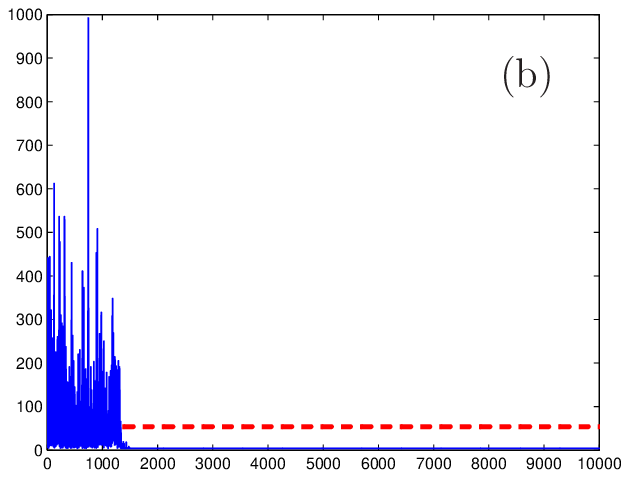}
\includegraphics[width=4.2cm]{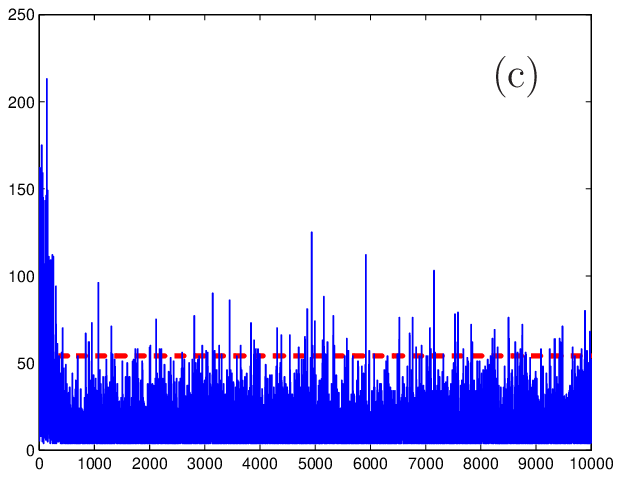}
\includegraphics[width=4.2cm]{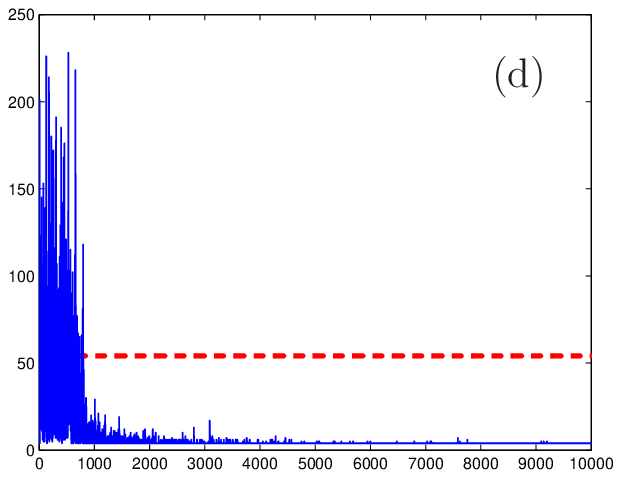}
\includegraphics[width=4.2cm]{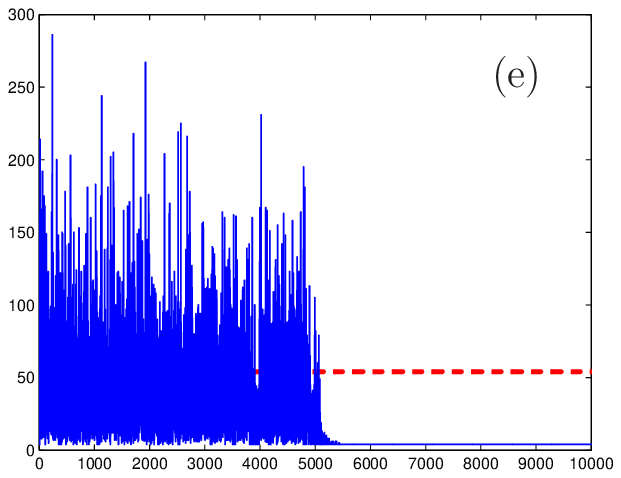}
\includegraphics[width=4.2cm]{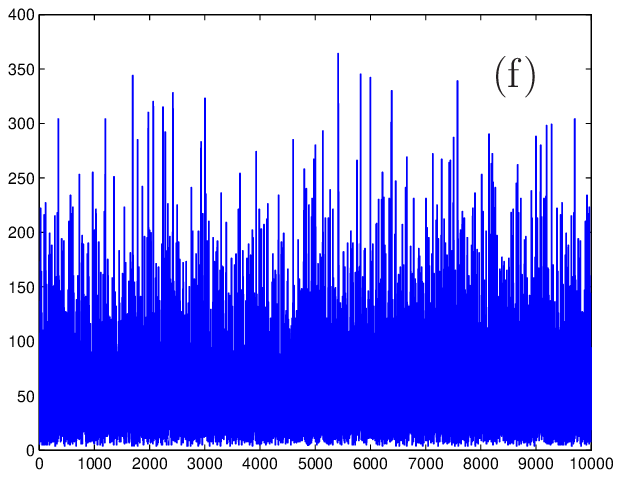}
\caption{\label{fig11}
Episode lengths as a function of the number of episodes for the
$(3\times3)$-grid world as shown in Fig.~\ref{fig10}. The plots illustrate the
effect of gradient glow for single histories of individual agents.
(a) The agent receives a reward $r$ $\!=$ $\!1$ when it has found the goal,
without gradient glow ($\eta$ $\!=$ $\!1$).
(b) As in (a), but with gradient glow enabled ($\eta$ $\!=$ $\!0.01$).
(c) In addition to receiving a reward $r$ $\!=$ $\!1$ when it has found the
goal, the agent is punished with a reward $r$ $\!=$ $\!-10$, when it has hit a
boundary, without gradient glow ($\eta$ $\!=$ $\!1$).
(d) As in (c), but with gradient glow enabled ($\eta$ $\!=$ $\!0.7$).
(e) As in (d), but with  gradient glow prolonged further
($\eta$ $\!=$ $\!0.5$).
(f) Learning is disabled by always setting the reward to 0. The agent performs a
random walk through the grid with average length 54.1 which is included as
dashed line in Figs.~\ref{fig11} and \ref{fig12}.
}
\end{figure}

\begin{figure}[ht]
\includegraphics[width=8.6cm]{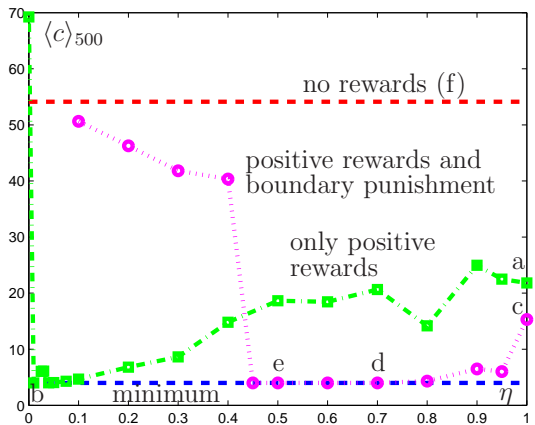}
\caption{\label{fig12}
Episode lengths $c$ averaged over the last 500 episodes in Fig.~\ref{fig11},
i.e., episodes $9.5*10^3$ $\!-$ $\!1*10^4$ of single histories of individual
agents in the $(3\times3)$-grid world as shown in Fig.~\ref{fig10}. The data
points distinguish various rewarding strategies and values of gradient glow
$\eta$ as explained in Fig.~\ref{fig11}. The upper and lower dashed lines
reflect a random walk and the shortest path, respectively, and the letters
(a)-(f) correspond to the respective plots in Fig.~\ref{fig11}. The optimum
value of $\eta$ depends on the details of the rewarding strategy.
}
\end{figure}

In Figs.~\ref{fig11} and \ref{fig12} we investigate the episode length which
is defined as the number of cycles per episode. Rather than performing an
ensemble average, we consider individual agents. These agents are described by
(\ref{ur3}) with a learning rate of $\alpha$ $\!=$ $\!10^{-1}$, absence of
relaxation ($\kappa$ $\!=$ $\!0$), and varying amounts of gradient glow
($\eta$ $\!\le$ $\!1$). The number of episodes equals the number of times the
agent is allowed to restart from S, whereas the time passed equals the sum of
episode lengths. The episode length can be infinite but not smaller than four,
the length of the shortest path from S to G.

Fig.~\ref{fig11} shows evolutions of episode lengths with the number of
episodes, where we have set a maximum of $10^4$ episodes. As explained, each
episode starts at S and ends only when G has been reached.
Fig.~\ref{fig11}(f) shows for comparison the lengths of $10^4$ random walks
through the grid of an agent whose learning has been disabled by always setting
the reward to 0. The average number of 54.1 steps to reach G from S is 
shown in Figs.~\ref{fig11} and \ref{fig12} as a dashed line for comparison.
In Figs.~\ref{fig11}(a-e), a positive reward of $r$ $\!=$ $\!1$ is given for
hitting G. While in Fig.~\ref{fig11}(a), the reward is always zero before G
has been hit, in Fig.~\ref{fig11}(c) hitting a boundary is punished with a
negative reward of $r$ $\!=$ $\!-10$, which slightly improves the agent's
performance. [Note that all plots are specific to the respective learning rate
(here $\alpha$ $\!=$ $\!10^{-1}$), which has been chosen by hand to observe an
improvement within our $10^4$ episode-window and at the same time
minimising the risk of oversized learning steps. While in general, the learning
rate is gradually decreased (cf. the conditions Eq. (2.8) in
\cite{bookSuttonBarto} to ensure convergence), this is not strictly necessary.
In our numerical examples we have kept $\alpha$ constant for simplicity.
Implementation of a dynamic adaptation of the learning rate as was done in
\cite{clausen18} and \cite{Cla15} in the present context is left for future
work.] The transitions Fig.~\ref{fig11}(a$\to$b) and
Fig.~\ref{fig11}(c$\to$d) show the effect of enabling gradient glow,
i.e. $(\eta=1)$ $\!\to$ $\!(\eta<1)$ in (\ref{ur3}). Gradient glow provides a
mechanism of gradual backpropagation of the policy change from the nearest
neighbourhood of G to cells more distant from G as the number of episodes
increases. In Fig.~\ref{fig11}, the agent settles in the optimal policy in
cases (b), (d) and (e).

The policy resulting after $\!10^{4}$ episodes in case Fig.~\ref{fig11}(d) is
given in Fig.~\ref{fig10}, where the numbers in each cell present the
probability to move in the respective direction. While the agent finds the
optimal policy for all cells forming the shortest path, it remains ignorant for
the remaining cells. As the agent finds and consolidates the shortest path, then
episode over episode, it soon visits the off-path cells less frequently, so that
the transition probabilities from these cells do not accumulate enough
iterations and are ``frozen'' in suboptimal values. This is characteristic of RL
and can also be observed in learning to play games such as Backgammon
\cite{bookSuttonBarto}, where it is sufficient to play well only in typical
rather than all possible constellations of the game. Since for large games, the
former often form a small subset of the latter, this can be seen as a strategy
to combat with large state spaces (such as number of possible game
constellations). To find an optimal policy for \emph{all} cells in
Fig.~\ref{fig10}, we may start each episode from a random cell, analogous to
initialising the agent in an overcomplete basis as explained in
Fig.~\ref{fig8}. The red numbers in parentheses shown in Fig.~\ref{fig10}
present a new policy obtained after $\!10^{5}$ episodes in this way. In contrast
to the old policy, it is optimal or nearly optimal for all cells, with the
difference between 1 and the sum of these numbers quantifying the deviation
from optimality for each cell $({\scriptstyle\stackrel{<}{=}}0.02)$. Since on
average, the agent starts from a given cell only in 1/7-th of all episodes, the
learning is slowed down, analogous to Fig.~\ref{fig8}(a).

Fig.~\ref{fig12} summarises the effect of gradient glow illustrated in
Fig.~\ref{fig11} for the two rewarding strategies.
To limit the numerical effort, we have averaged the episode lengths over the
last 500 episodes in Fig.~\ref{fig11} for individual agents as a ``rule
of thumb''-measure of the agent's performance for the strategy chosen. For a
deterministic calculation we must instead average the length of each episode
(and for each $\eta$) over a sufficiently large ensemble of independent agents
for as many episodes as needed to reach convergence. Despite these shortcomings,
the results indicate a qualitatively similar behaviour as Fig. 4(a) in
\cite{Mel14}. Figs.~\ref{fig11} and \ref{fig12} demonstrate that gradient
glow improves the agent performance, irrespective of whether or not it
receives information on false intermediate moves by means of negative rewards,
although the latter reduce the required length of glow.
It is expected that for an ensemble average, an optimal value of $\eta$ can be
found, with which the fastest convergence to the shortest path can be achieved.
Fig.~\ref{fig11} distinguishes two qualitatively different modes of
convergence. If $\eta$ is larger than optimal, a gradual improvement is
observed, as seen by the damping of spikes in Fig.~\ref{fig11}(d). If $\eta$
is smaller than optimal, then an abrupt collapse to the optimal policy without
visible evidence in the preceding statistics that would provide an indication is
observed, cf. Fig.~\ref{fig11}(e). If $\eta$ is decreased further, this
transition is likely to happen later, to the point it will not be observed
within a fixed number of episodes. This results in the steep increase in episode
length shown in Fig.~\ref{fig12}, which would be absent if the ensemble
average was used instead. This sudden transition as shown in
Fig.~\ref{fig11}(e) can also be observed for individual agents in \cite{Mel14}
(not shown there), which applies a softmax-policy function along with edge glow.
It is surprising that the quadratic measurement-based policy simulated here
exhibits the same phenomenon. Note however, that convergence does not imply
optimality. In tabular RL and PS, such an abrupt transition can be observed if
the $\lambda$-parameter and hence the ``correlation length'' is too large
(in RL) or if the $\eta$-parameter is too small, so that the glow lasts too long
(in PS). The policies obtained in this way are typically sub-optimal, especially
in larger scale tasks such as bigger grid worlds, for which the agent learns
``fast but bad'' in this case. It is hence expected that a similar behaviour can
be observed for our method if we increased the size of the grid.
\section{
\label{sec6}
Finite difference updates}
This work's numerical experiments rely on a symbolic expression
(\ref{gradcomps}) for the gradient $\bm{\nabla}_{t}$ in (\ref{grad}) for
simplicity, which is usually not available in practice, also keeping in mind the
variety of compositions Fig.~\ref{fig2}, so that the agent's memory
$\hat{U}(\bm{h})$ is generally unknown. As explained in the discussion of
Fig.~\ref{fig1}, the agent may then apply a measurement-based internal loop by
repeatedly preparing its memory in a state that corresponds to the last percept
$s_t$, and register whether or how often the last measurement outcome $a_t$ can
be recovered. This approach can be done with either infinitesimal or finite
changes in the control vector $\bm{h}$, where we can distinguish between
expectation value- and sample-based updates, depending on how many internal
cycles are performed between consecutive external cycles.
It should be stressed that the external cycles in Fig.~\ref{fig1}
represent the agent-environment interaction, resulting in sequences of
state-action pairs and corresponding rewards. While in an elementary optimal
control problem, a given objective is to be optimized, here the environment
poses at each external cycle a separate and generally unpredictable control
problem, all of which must be addressed by the agent simultaneously.

\begin{figure}[ht]
\includegraphics[width=7cm]{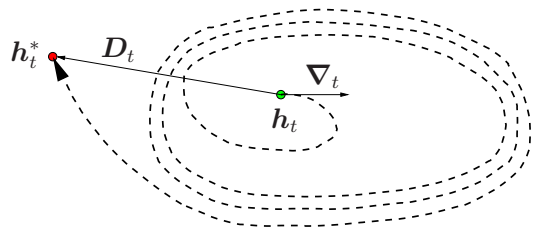}
\caption{\label{fig13}
Following the direction $\bm{\nabla}p(a_t|s_t)$ of steepest ascent (dashed line)
does not necessarily lead to the shortest route
$\bm{D}_t$ $\!=$ $\!\bm{h}_t^*$ $\!-$ $\!\bm{h}_t$
from a given control vector $\bm{h}_t$ at cycle $t$ to a location $\bm{h}_t^*$,
for which $p(a_t|s_t)$ becomes maximum.
}
\end{figure}

Due to the small learning rate $\alpha$, the update rule (\ref{ur3}) is in all
cases local in parameter space, which reflects the assumption, that a physical
agent cannot completely reconfigure its ``hardware'' in a single instant. While
it is then consistent to apply a gradient $\bm{D}_t$ $\!=$ $\!\bm{\nabla}_t$ as
a local quantity in (\ref{ur3}), from a computational perspective, it has a few
drawbacks, however. One is that the direction of steepest accent at the current
control vector $\bm{h}_t$ does not need to coincide with the direction
$\bm{D}_t$ $\!=$ $\!\bm{h}_t^*$ $\!-$ $\!\bm{h}_t$ towards the optimum
$\bm{h}_t^*$, as illustrated in Fig.~\ref{fig13}.

Another aspect is the vanishing of the gradient. Consider for example the
initialisation of the action system in a mixed state (\ref{pcoh}) as done in
Fig.~\ref{fig5}(b). In particular, the graph with
$p_{\mathrm{coh}}$ $\!=$ $\!0$ does not display any learning ability.
Substituting the corresponding
$\hat{\varrho}_{\mathrm{A}}$ $\!=$ $\!\hat{I}_{\mathrm{A}}/2$ in (\ref{pcoh})
and $\hat{U}$ $\!=$ $\!\hat{I}$ into (\ref{gradcomps}), we see that the reason
is the vanishing gradient, $\nabla_k$ $\!=$
$\!\mathrm{Im}\mathrm{Tr}[\hat{\varrho}_{\mathrm{S}}\hat{\Pi}({a})\hat{H}_k]$
$\!=$ $\!0$. On the other hand, the corresponding setup Fig.~\ref{fig6}(a)
reveals that in this case, substituting a SWAP-gate between S and A for
$\hat{U}$ provides an optimal solution (along with an X-gate if the meaning of
the symbols is reversed) for any $\hat{\varrho}_{\mathrm{A}}$, that is obviously
not found in Fig.~\ref{fig5}(b). This failure occurs despite the fact that the
agents explore, as indicated by the fluctuations in Fig.~\ref{fig5}(b). To
understand the difference, note that we may generate an $\varepsilon$-greedy
policy function by replacing in (\ref{grada}) an (arbitrarily given) state
$\hat{\varrho}({s})$ with
$\frac{\hat{\varrho}({s})+\varepsilon\hat{I}}{1+\varepsilon{d}}$,
where $0$ $\!<$ $\!\varepsilon$ $\!\ll$ $\!1$ and 
$d$ $\!=$ $\!\mathrm{Tr}\hat{I}$. The term with $\hat{I}$ then gives to
(\ref{grada}) a contribution $\sim\mathrm{Tr}_{\mathrm{A}}\hat{\Pi}({a})$, that
is independent of $s$. At the same time, it does not contribute in
(\ref{gradcomps}) to the gradient, $\nabla_k$ $\!=$ $\!0$.
If $\bm{D}_t$ $\!=$ $\!\bm{\nabla}_t$ $\!=$ $\!0$ for all $t$ in (\ref{ur3}),
the agent's learning comes to rest, however. Finite difference and
sample-based updates here offer a possibility to explore
\emph{in parameter space} the neighbourhood of the present location $\bm{h}_t$
(or, colloquially, the ``state'') of the agent's memory, as a consequence of
asymmetries in the control landscape or statistical fluctuations in the samples.

Of particular relevance is a final fixpoint (\ref{task}). Intuitively, one
would assume that (despite the compactness of the (S)U($n$)-groups, that is in
contrast to the potentially unbounded values of $U$ in RL or $h$ in PS) once an
agent has settled in a point (\ref{task}), due to the vanishing gradient, it
won't be able to react quickly, if the environment suddenly changes its
allocation of rewards (without confronting the agent with percepts it has not
perceived before). However, the learning curves for controllable memories (16
and 32 controls) in Fig.~\ref{fig7}(a) demonstrate that relearning after
$5\cdot10^3$ cycles is not affected. A study of individual agents with 32
controls in Fig.~\ref{fig7}(a) reveals that the Euclidean length of the
numerical gradient rises from $10^{-14}$ at cycle 5000 to a value $>1$ in only
15 cycles. Better understanding of this is left for future study.
In what follows, we outline the mentioned alternatives in some more detail.
\subsection{
Expectation value-based updates}
If the time consumed by the internal cycles is uncritical with respect to the
external cycles, the agent can obtain estimates of $p(a_t|s_t)$ from a
sufficiently large number of internal binary measurements. With these, it can
either approximate the components $\nabla_kp(a_t|s_t;{h}_j)$ $\!\approx$
$\![p(a_t|s_t;{h}_j+\delta_{jk}\delta{h}_k)$ $\!-$
$\!p(a_t|s_t;{h}_j)]/\delta{h}_k$ of the local gradient
$\bm{\nabla}_t$ $\!=$ $\!\bm{\nabla}p(a_t|s_t;\bm{h}_t)$, which is then
substituted as $\bm{D}_t$ $\!=\bm{\nabla}_t$ into (\ref{ur3}). Alternatively, it
can perform a global search for the location $\bm{h}_t^*$ of the maximum of
$p(a_t|s_t)$. A possible algorithm for the latter is differential evolution,
which relies on deterministic values $p(a_t|s_t;\bm{h})$ rather than noisy
samples. Once an estimate for $\bm{h}_t^*$ has been found, the difference
$\bm{D}_t$ $\!=$ $\!\bm{h}_t^*$ $\!-$ $\!\bm{h}_t$ is used in (\ref{ur3}).
\subsection{
Sample-based updates}
Reliance on expectation values may give away potential speed gains offered by a
quantum memory, which poses the question, whether a finite number of sample
measurements is sufficient. Since individual updates in (\ref{ur3}) are made
with a small fraction $\alpha$ of the whole $\bm{D}_t$, the assumption is that
the individual statistical errors in the sampled $\bm{D}_t$ cancel out in the
long run.

\begin{figure}[ht]
\includegraphics[width=5cm]{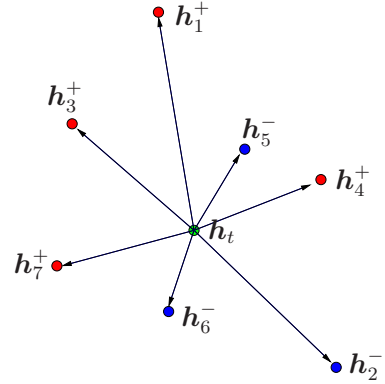}
\caption{\label{fig14}
Internally generated random cloud of sample controls $\bm{h}_k$ around
a given control vector $\bm{h}_t$ at cycle $t$ for which binary measurements
``given $s_t$, detect $a_t$ or not'' are carried out between external cycles,
yielding positive ($\bm{h}_k^+$) or negative ($\bm{h}_k^-$) outcomes.
}
\end{figure}

As for the expectation value-based updates discussed above, samples can be used
to either create \emph{discrete} estimates
$\nabla_kp(a_t|s_t;{h}_j)$ $\!\approx$ $\![s({h}_j+\delta_{jk}\delta{h}_k)$
$\!-$ $\!s({h}_j)]/2$ for the components $k$ of the local gradient
$\bm{\nabla}_t$ $\!=$ $\!\bm{\nabla}p(a_t|s_t;\bm{h}_t)$, where
$s$ $\!=$ $\!s(\bm{h}_t)$ $\!=$ $\!\pm1$ depending on whether the outcome of the
binary measurement $(a_t|s_t;\bm{h}_t)$ is positive or not. Alternatively, for
finite difference updates, one may consider a neural gas
\cite{MaSc91,*Fritzke95agrowing} inspired approach depicted in Fig.~\ref{fig14}.
In this approach, the differences
\begin{equation}
\label{ng}
  \bm{D}_t^{(n)}=\frac{1}{n}\sum_{k=1}^ns_k\bm{h}_k
  =\frac{n-1}{n}\bm{D}_t^{(n-1)}+\frac{1}{n}s_n\bm{h}_n
\end{equation}
between the sampled centers of positive
[$s_k$ $\!=$ $\!s(\bm{h}_k)$ $\!=$ $\!+1$, i.e.,
$\bm{h}_k$ $\!=$ $\!\bm{h}_k^+$] and negative outcomes
($s_k$ $\!=$ $\!s(\bm{h}_k)$  $\!=$ $\!-1$, i.e.,
$\bm{h}_k$ $\!=$ $\!\bm{h}_k^-$) of the binary measurements $(a_t|s_t;\bm{h}_k)$
are then applied in (\ref{ur3}). Although one could store a current estimate
$\bm{D}_t^{(n)}$ for each observed state-action pair $(a_t|s_t)$ and merely
update it according to (\ref{ng}) with each new measurement point
$s_{n+1}\bm{h}_{n+1}$, this would give away the generalization capability
described in Sec.~\ref{sec:nec}. One would hence need to perform $n$ internal
cycles with the POVM-based internal loop between each external cycle.
The $\bm{h}_k$ could be drawn, e.g., from a Gaussian centered around the
respective $\bm{h}_t$. The variance of this Gaussian could be gradually
decreased with the number of external cycles to increase the locality (local
resolution) of the cloud.

Fig.~\ref{fig13} gives the misleading impression that finite difference updates
are superior to gradient-based methods. To give an illustrative counterexample,
one could think of a two-dimensional $\bm{h}$ $\!=$ $\!(h_x,h_y)$ and a control
landscape $p(\bm{h})$ modelled by the monotonically increasing height $p(l)$
along the length $l$ of a tape of paper bent into a spiral and placed onto the
dashed line in Fig.~\ref{fig13}, such that one end with $p(0)$ $\!=$ $\!0$ is
located at $\bm{h}_t$ and the other one at $\bm{h}_t^*$. Here, a gradient-based
method would safely follow the long path on the tape's upper edge, whereas a
finite difference method would trade a potential speedup with the risk of
missing the paper at all trials. Since a comparison of state of the art optimal
control methods based on noisy samples for the agent's learning would go beyond
the scope of this work, we here restrict ourselves to these sketchy lines of
thought, whose numerical study is pending, and leave open the question of what
the best method is for a given task.

A characteristic shared by the loops of Fig.~\ref{fig1} and
optimal control setups is the need of an experimental ``mastermind'' who
controls the controls. An agent which is supposed to act autonomously would be
required to accomplish this by itself, ideally in a ``natural'' or ``organic
computing'' sense. An elementary example from everyday life are ``desire paths''
which form or dissipate, depending on their usage and without a designated
planner.
\section{
\label{sec5}
Summary and outlook}
In summary, we have adopted an update rule from the basic PS scheme, equipped it
with gradient glow, and applied it to small-scale invasion game and grid
world tasks. The numerical results show that similar results can be obtained for
a quantum agent, as long as the memory is not underactuated. This is not
obvious, because of the fundamental difference in the number of free parameters.
If $S$ and $A$ denote the number of possible percepts and actions, respectively,
then in classical tabular action value RL-methods, the estimated values of all
percept-action pairs are combined to a $({S}\times{A})$-matrix, i.e., we have
$(SA)$ real parameters. If we encoded in our scheme each percept and action
category by a separate subsystem, whose dimensionalities correspond to the
number of values, the respective category can adopt, then $\hat{U}$ is an at
least $U(N=SA)$-matrix for which we are faced with $(SA)^2$ real parameters.
Note that this work is unrelated to the reflecting PS agents, which are
discussed in \cite{Pap14}. While the scheme \cite{Pap14} allows a proof of
quantum speedup, our approach complements the latter in that it is simple,
flexible in its construction, and does not involve specific analytic quantum
gates. The learning of a good policy only for percepts which are ``typical'' and
have thus been encountered sufficiently often in the past shares features with
``soft computing'', where it is sufficient to find a good rather than an exact
solution, which would here consist in a policy that is optimal for all possible
percepts. One may think of, e.g., simplifying a symbolic mathematical
expression: while all transformation steps themselves must be exact, there are
no strict rules, as far as the best way of its formulation is concerned. In
future work, it may be worth to incorporate recent extensions of the classical
PS scheme such as generalization \cite{Mel15}.
\appendix
\section{
\label{app:A1}
Classical PS update rule}
Classical PS in its general form is based on a discrete network of clips that
form its ECM. In the examples considered in \cite{Bri12,Mau15,Mel14}, after each
discrete time step (external cycle), a local edge is updated according to
\begin{equation}
  h_{t+1}=h_t-\gamma(h_t-h^{\mathrm{eq}})+r_t,
\end{equation}
where the instantaneous excitation $r_t$ $\!=$ $\!g_t\lambda_t$ is the product
of the edge's current glow value $g_t$ and the respective non-negative reward
$\lambda_t$ given at this time step. The glow values dampen according to
$g_{t+1}$ $\!=$ $\!(1-\eta)g_t$ with $g_0$ $\!=$ $\!0$, where $\eta\in[0,1]$ is
a glow parameter, and are reset to 1 if the edge was visited during a cycle.
$\gamma\in[0,1]$ is a damping parameter towards an equilibrium value
$h^{\mathrm{eq}}$ in the absence of rewards (e.g., 1). Starting from some given
$h_0$ (e.g., $h^{\mathrm{eq}}$), this gives
\begin{eqnarray}
  h_t&=&(1-\gamma)^th_0+[1-(1-\gamma)^t]h^{\mathrm{eq}}
  +\sum_{k=0}^{t-1}(1-\gamma)^kr_{t-k-1}
  \nonumber\\
  &=&h_0+\sum_{k=0}^{t-1}r_{t-k-1}\quad(\gamma=0),
\label{hdiv}
  \\
  h_t&\approx&h^{\mathrm{eq}}+\sum_{k=0}^{t-1}(1-\gamma)^kr_{t-k-1}
  \quad(\gamma>0),
\label{happ}
\end{eqnarray}
where the approximation as a backward-discounted sum (\ref{happ}) holds for
times sufficiently large so that $(1-\gamma)^t$ $\!\ll$ $\!1$ or exactly for
$h_0$ $\!=$ $\!h^{\mathrm{eq}}$. Note that due to the absence of damping,
(\ref{hdiv}) diverges in general if $t$ grows without limit.
\section{
\label{app:A2}
Updating the memory $\hat{U}$}
In App.~\ref{app:A2} we focus on a model-based (i.e., symbolic) determination
of the gradient $\bm{\nabla}\hat{U}(\bm{h})$. The exact form of the gradient
depends on the parametrization. For example, if
$\hat{U}(\bm{h})$ $\!=$ $\!\mathrm{e}^{-\mathrm{i}\hat{H}(\bm{h})}$ is given by
some Hermitian $\hat{H}$, then
\begin{equation}
\label{x1}
  \bm{\nabla}\hat{U}=\int_0^1
  \mathrm{e}^{-\mathrm{i}x\hat{H}}(-\mathrm{i}\bm{\nabla}\hat{H})
  \mathrm{e}^{-\mathrm{i}(1-x)\hat{H}}\mathrm{d}x.
\end{equation}
For small $\hat{H}$, we can expand the exponentials in $\hat{H}$ to
lowest order, and the approximation
$\bm{\nabla}\hat{U}$ $\!\approx$ $\!-\mathrm{i}\bm{\nabla}\hat{H}$ holds.

In case of a continuous time dependence, the vector $\bm{h}$ can be replaced by
a function $h(t)$, with which a unitary propagator from time $t_1$ to time $t_3$
is given as a positively time ordered integral
\begin{equation}
\label{x2}
  \hat{U}(t_3,t_1)=\mathrm{T}\mathrm{e}^{-\mathrm{i}
  \int_{t_1}^{t_3}\mathrm{d}t_2h(t_2)\hat{H}(t_2)},
\end{equation}
and
\begin{equation}
\label{x3}
  \frac{\delta\hat{U}(t,0)}{\delta{h}}(t_1)
  =-\mathrm{i}\hat{U}(t,t_1)\hat{H}(t_1)\hat{U}(t_1,0).
\end{equation}
If $\hat{H}(t)$ $\!=$ $\!\sum_k\tilde{h}_k(t)\hat{H}_k$ is expanded in terms of
some fixed Hamiltonians $\hat{H}_k$, then with $h_k$ $\!=$ $\!h\tilde{h}_k$,
(\ref{x2}) becomes
$\hat{U}(t_3,t_1)$ $\!=$ $\!\mathrm{T}\mathrm{e}^{-\mathrm{i}
\int_{t_1}^{t_3}\mathrm{d}t_2\sum_kh_k(t_2)\hat{H}_k}$, and (\ref{x3}) is
replaced with $\frac{\delta\hat{U}(t,0)}{\delta{h_k}}(t_1)$ $\!=$
$\!-\mathrm{i}\hat{U}(t,t_1)\hat{H}_k\hat{U}(t_1,0)$.
Navigation on unitary groups becomes discretized if only a restricted (finite)
set of Hamiltonians $\hat{H}_k$ can be implemented at a time rather than an
analytically time-dependent Hamiltonian $\hat{H}(t)$, so that only one of the
$\tilde{h}_k$ is non-zero for a given $t$.
A known example is the alternating application of two fixed Hamiltonians
\cite{lloyd2}, $\hat{H}_{2k}$ $\!=$ $\!\hat{H}^{(2)}$ and
$\hat{H}_{2k+1}$ $\!=$
$\!\hat{H}^{(1)}$ $(k=0,1,2,\ldots,\lfloor\frac{n}{2}\rfloor)$, for a set of
times to be determined from the target unitary \cite{akulin2}.
In this discrete case as defined by a piecewise constant normalized $\hat{H}$
in (\ref{x2}), the function $h(t)$ can be replaced with a vector $\bm{h}$, and
the functional derivatives with respect to $h(t)$ reduce to gradients with
respect to $\bm{h}$.
\subsection{
\label{app:lxl}
Adding layer after layer}
We can update the present unitary $\hat{U}$ by multiplying it from the left with
a new layer $\hat{U}(\delta\bm{h})$ after each cycle,
\begin{equation}
\hat{U}\leftarrow\hat{U}(\delta\bm{h})\hat{U}.
\end{equation}
If $\hat{U}(\delta\bm{h})$ $\!=$
$\!\mathrm{e}^{-\mathrm{i}\sum_k\delta{h}_k\hat{H}_k}$ is a small modification
close to the identity, then the mentioned approximation of (\ref{x1}) gives
$(\bm{\nabla}\hat{U})_k$ $\!\approx$ $\!-\mathrm{i}\hat{H}_k\hat{U}$. This
is of advantage if the agent or its simulation is able to store only the present
$\hat{U}$ and not the history of updates (layers). The components of
(\ref{grad}) then become
\begin{equation}
  \frac{\partial{p}({a}|{s})}{\partial{h}_k}
  =2\mathrm{Im}
  \bigl\langle\hat{U}^\dagger\hat{\Pi}({a})\hat{H}_k\hat{U}\bigr\rangle.
\end{equation}
\subsection{
\label{app:fl}
Fixed number of layers}
In our numerical examples we consider a discretized memory model, for which
(\ref{x2}) reduces to a product of $n$ unitaries
\begin{equation}
  \hat{U}=\hat{U}_n\cdots\hat{U}_2\hat{U}_1,\quad
  \hat{U}_k=\mathrm{e}^{-\mathrm{i}{h}_k\hat{H}_k},
\end{equation}
which simplifies the determination of the gradient, since
$(\bm{\nabla}\hat{U})_k$ $\!=$ $\!-\mathrm{i}\hat{U}\hat{H}_k(t_k)$, where
$\hat{H}_k(t_k)$ $\!=$
$\!(\hat{U}_k\cdots\hat{U}_1)^\dagger\hat{H}_k(\hat{U}_k\cdots\hat{U}_1)$, so
that the components of (\ref{grad}) now become
\begin{equation}
\label{gradcomps}
  \frac{\partial{p}({a}|{s})}{\partial{h}_k}
  =2\mathrm{Im}
  \bigl\langle\hat{U}^\dagger\hat{\Pi}({a})\hat{U}\hat{H}_k(t_k)
  \bigr\rangle.
\end{equation}
In this work, we use alternating layers defined by two fixed Hamiltonians
$\hat{H}^{(1)}$ and $\hat{H}^{(2)}$ as mentioned at the beginning of this
section.
\section{Numerical and measurement-based objectives}
\subsection{Distance and uniformly averaged fidelity}
Consider an $n$-level system and two unitary operators $\hat{U}$ and
$\hat{U}_{\mathrm{T}}$, where $\hat{U}$ $\!=$ $\!\hat{U}({h})$ depends on an
(unrestricted) external control field $h(t)$, and $\!\hat{U}_{\mathrm{T}}$ is a
desired target. Their (squared) Euclidean distance as induced by the
Hilbert-Schmidt dot product is given by
\begin{eqnarray}
  D&\equiv&\|\hat{U}-\hat{U}_{\mathrm{T}}\|^2
  =\mathrm{Tr}\bigl[(\hat{U}-\hat{U}_{\mathrm{T}})^\dagger
  (\hat{U}-\hat{U}_{\mathrm{T}})\bigr]
  \nonumber\\
  &=&2n-2\mathrm{Re}\mathrm{Tr}(\hat{U}_{\mathrm{T}}^\dagger\hat{U})
  \in[0,4n].
\label{D}
\end{eqnarray}
If $\hat{U}({h})$ is controllable in the sense that at least one $h(t)$ exists
such that $\hat{U}({h})$ $\!=$ $\!\hat{U}_{\mathrm{T}}$, then (\ref{D}) has the
set $\{0,4,\ldots,4n\}$ as possible extremal values
(i.e, $\frac{\delta{D}}{\delta{h}}$ $\!=$ $\!0$), where the values $0$ and $4n$
are attained for $\hat{U}$ $\!=$ $\!\pm\hat{U}_{\mathrm{T}}$, while the
remaining extrema are saddle points \cite{Rab05}. A measure insensitive to
global phases
$\hat{U}$ $\!=$ $\!\mathrm{e}^{\mathrm{i}\varphi}\hat{U}_{\mathrm{T}}$ is the
average fidelity defined by
\begin{equation}
\label{F}
  F\equiv
  \overline{|\langle\Psi|\hat{U}_{\mathrm{T}}^\dagger\hat{U}|\Psi\rangle|^2}
  =\frac{n+|\mathrm{Tr}(\hat{U}_{\mathrm{T}}^\dagger\hat{U})|^2}{n(n+1)},
\end{equation}
where the overline denotes uniform average over all $|\Psi\rangle$
\cite{Ped07,dan05}. Note that $F$ $\!\in$ $\![(n+1)^{-1},1]$ for
$n$ $\!>$ $\!1$, and $F$ $\!=$ $\!1$ for $n$ $\!=$ $\!1$.
Both (\ref{D}) and (\ref{F}) are determined by the complex
\begin{equation}
  \cos\sphericalangle\bigl(\hat{U},\hat{U}_{\mathrm{T}}\bigr)
  \equiv\frac{\mathrm{Tr}(\hat{U}_{\mathrm{T}}^\dagger\hat{U})}
  {\sqrt{\mathrm{Tr}(\hat{U}^\dagger\hat{U})}
  \sqrt{\mathrm{Tr}(\hat{U}_{\mathrm{T}}^\dagger\hat{U}_{\mathrm{T}})}}
  =\frac{\mathrm{Tr}(\hat{U}_{\mathrm{T}}^\dagger\hat{U})}{n},
\end{equation}
which is confined to the complex unit circle and whose expectation
$\bigl\langle|\cos\sphericalangle(\hat{U},\hat{U}_{\mathrm{T}})|^2\bigr\rangle$
$\!=$ $\!n^{-1}$ for uniform random $\hat{U}$ drops to zero with growing $n$.

If in (\ref{F}), the averaging is restricted to a $d$-dimensional subspace P,
we must replace (\ref{F}) with
\begin{equation}
\label{FP}
  F_{\mathrm{P}}
  \equiv\overline{|\langle\Psi|\hat{M}|\Psi\rangle|^2}^{(\mathrm{P})}
  =\frac{\mathrm{Tr}(\hat{M}^\dagger\hat{M})+|\mathrm{Tr}(\hat{M})|^2}{d(d+1)},
\end{equation}
where $\hat{M}$ $\!=$ $\!\hat{P}\hat{U}_{\mathrm{T}}^\dagger\hat{U}\hat{P}$,
with $\hat{P}$ being the projector onto P. Note that
$F_{\mathrm{P}}$ $\!\in$ $\![\frac{\max(0,2d-n)}{d(d+1)},1]$ for
$n$ $\!>$ $\!1$ [since $\mathrm{Tr}(\hat{M}^\dagger\hat{M})$ $\!=$
$\!\mathrm{Tr}(\hat{P}_{\mathrm{T}}\hat{P}_{\mathrm{U}})$ $\!\in$
$\!\max(0,2d-n)$ with $\hat{P}_{\mathrm{T}}$ $\!=$
$\!\hat{U}_{\mathrm{T}}\hat{P}\hat{U}_{\mathrm{T}}^\dagger$,
$\hat{P}_{\mathrm{U}}$ $\!=$ $\!\hat{U}\hat{P}\hat{U}^\dagger$], 
and $F_{\mathrm{P}}$ $\!=$ $\!1$ for $n$ $\!=$ $\!1$.
While for a one-dimensional $\hat{P}$ $\!=$ $\!|\Psi\rangle\langle\Psi|$,
(\ref{FP}) reduces to $F_\Psi$ $\!=$
$\!|\langle\Psi|\hat{U}_{\mathrm{T}}^\dagger\hat{U}|\Psi\rangle|^2$, the other
limit $d$ $\!=$ $\!n$ recovers (\ref{F}).

If in (\ref{FP}), $\hat{U}$ $\!=$ $\!\hat{U}_{\mathrm{SB}}$ couples the quantum
system S to a bath B, then we define a projector
$\hat{\Pi}$ $\!=$ $\!\hat{U}_{\mathrm{T}}\hat{P}|\Psi\rangle\langle\Psi|
\hat{P}\hat{U}_{\mathrm{T}}^\dagger\otimes\hat{I}_{\mathrm{B}}$ and generalize
(\ref{FP}) to
\begin{eqnarray}
\label{FPB}
  F_{\mathrm{P}}
  &\equiv&\overline{\mathrm{Tr}_{\mathrm{SB}}\bigl[\hat{U}\hat{P}|\Psi\rangle
  \hat{\varrho}_{\mathrm{B}}
  \langle\Psi|\hat{P}\hat{U}^\dagger\hat{\Pi}\bigr]}^{(\mathrm{P})}
  \\
  &=&\left\langle
  \frac{\mathrm{Tr}_{\mathrm{S}}(\hat{M}^\dagger\hat{M})
  +(\mathrm{Tr}_{\mathrm{S}}\hat{M})^\dagger(\mathrm{Tr}_{\mathrm{S}}\hat{M})
  }{d(d+1)}\right\rangle_{\mathrm{B}},
\end{eqnarray}
where $\langle\cdots\rangle_{\mathrm{B}}$ $\!\equiv$
$\!\mathrm{Tr}_{\mathrm{B}}[\hat{\varrho}_{\mathrm{B}}(\cdots)]$ with a fixed
bath state $\hat{\varrho}_{\mathrm{B}}$.

Replacing $\hat{U}|\Psi\rangle\langle\Psi|\hat{U}^\dagger$ in (\ref{F}) with the
output $\mathcal{M}(|\Psi\rangle\langle\Psi|)$ of a quantum channel generalizes
(\ref{F}) to \cite{Ped07}
\begin{equation}
\label{F1}
  F=\overline{\langle\Psi|\hat{U}_{\mathrm{T}}^\dagger
  \mathcal{M}(|\Psi\rangle\langle\Psi|)
  \hat{U}_{\mathrm{T}}|\Psi\rangle}
  =\frac{n+\sum_k|\mathrm{Tr}(\hat{U}_{\mathrm{T}}^\dagger\hat{G}_k)|^2}
  {n(n+1)},
\end{equation}
where $\hat{G}_k$ are the Kraus operators of the decomposition of the channel
map $\mathcal{M}(\hat{\varrho})$ $\!=$
$\!\sum_k\hat{G}_k\hat{\varrho}\hat{G}_k^\dagger$. Note that a change
$\hat{G}_k^\prime$ $\!=$ $\!\sum_jV_{kj}\hat{G}_j$ of the Kraus operators as
described by a unitary matrix $V$ leaves (\ref{F1}) invariant.
\subsection{Percept statistics-based fidelity}
The uniform average in (\ref{F1}) can be generalized to an arbitrary
distribution ${p}(|\Psi_k\rangle)$ of possible input states $|\Psi_k\rangle$,
\begin{equation}
\label{rhoin}
  \hat{\varrho}_{\mathrm{in}}
  =\sum_k{p}(|\Psi_k\rangle)|\Psi_k\rangle\langle\Psi_k|,
\end{equation}
that reflects the statistics of their occurrence in different instances of
applications of the device (external cycles in a control loop). This generalizes
(\ref{F1}) to
\begin{eqnarray}
\label{F2}
  &&F_{\hat{\varrho}_{\mathrm{in}}}\equiv{p}(\hat{U}_{\mathrm{T}})
  =\sum_k{p}(\hat{U}_{\mathrm{T}}||\Psi_k\rangle){p}(|\Psi_k\rangle),
  \\
  &&{p}(\hat{U}_{\mathrm{T}}||\Psi_k\rangle)
  =\langle\Psi_k|\hat{U}_{\mathrm{T}}^\dagger
  \mathcal{M}(|\Psi_k\rangle\langle\Psi_k|)
  \hat{U}_{\mathrm{T}}|\Psi_k\rangle,\quad
\end{eqnarray}
which is just the total probability ${p}(\hat{U}_{\mathrm{T}})$ of correctly
detecting a $\hat{U}_{\mathrm{T}}$-transformed pure (but unknown) input state
drawn from a distribution (\ref{rhoin}).
Once $p(\hat{U}_{\mathrm{T}})$ $\!=$ $\!1$, the channel's effect is
indistinguishable from that of $\hat{U}_{\mathrm{T}}$ for the set of possible
inputs $|\Psi_k\rangle$, (i.e. those for which
${p}(|\Psi_k\rangle)$ $\!>$ $\!0$). The case
$p(\hat{U}_{\mathrm{T}})$ $\!\scriptstyle{\stackrel{<}{\approx}}$ $\!1$ is
relevant from a numerical and experimental point of view. Rare inputs
$|\Psi_k\rangle$, for which $0$ $\!<$ $\!{p}(|\Psi_k\rangle)$ $\!\ll$ $\!1$,
will hardly affect $p(\hat{U}_{\mathrm{T}})$ in a control loop, which relaxes
the demands on the channel compared to the uniform average (\ref{F1}). The
channel optimization itself is thus economized in the sense that it is required
to perform well only on \emph{typical} rather than \emph{all} inputs.

Here, we consider a navigation of $\hat{U}$ in the above-mentioned discretized
case, $\hat{U}$ $\!=$ $\!\hat{U}(\bm{h})$, that starts at the identity
$\hat{U}(\bm{h}$ $\!=$ $\!\bm{0})$ $\!=$ $\!\hat{I}$ and from there undertakes a
gradient-based maximization of ${p}(\hat{U}_{\mathrm{T}})$ as defined in
(\ref{F2}) to a point where
\begin{equation}
\label{task}
  \bm{\nabla}{p}(\hat{U}_{\mathrm{T}})=0.
\end{equation}
The control vector $\bm{h}$ typically represents system variables and not those
of the bath. Rather than solving for the parameters \cite{akulin2} for which
the scheme \cite{lloyd2} yields a desired given unitary, we search parameters,
for which the unitary, whose specific form we are not interested in, solves a
given task.
\subsection{Optimal memory navigation and constrained optimization}
While here we have discussed and compared concrete types of algorithms, a more
fundamental question concerns the optimality and derivation of general
(e.g., speed) limits of the learning process. Although the physical time is
given as the sum of the agent and environment response times over each cycle,
one may restrict to counting the number of (a) memory cycles in total, (b)
external cycles only (c) episodes or (d) parameter updates $\delta\bm{h}$ of the
memory $\hat{U}(\bm{h})$, depending on what the most critical criterion is. Not
only should the navigation of $\hat{U}$ from the identity to a point
(\ref{task}) follow an optimal trajectory, but also the navigation of the
percept states by a given $\hat{U}$ should be such that the individual physical
memory evolution times become minimum. Such demands may conflict with
restrictions on the practical implementability and complexity of the memory.
Since these questions are beyond the scope of this work, in what follows we
restrict ourselves to outline a connection to constrained optimisation as a
possible formal approach.

Assuming the Schr{\"o}dinger equation
{\small $\frac{\mathrm{d}}{\mathrm{d}t}\hat{U}$ $\!=$
$\!-\mathrm{i}\hat{H}\hat{U}$} and a fixed energy-type constraint
$\|\frac{\mathrm{d}}{\mathrm{d}t}\hat{U}\|^2$ $\!=$
$\!\mathrm{Tr}(\hat{H}^2)$ $\!\scriptstyle{\stackrel{!}{=}}$ $\!E^2$, the
length of a curve in the space of unitaries becomes
\begin{equation}
\label{L}
  L=\int_0^T\Bigl\|\frac{\mathrm{d}\hat{U}}{\mathrm{d}t}\Bigr\|\mathrm{d}t
  \stackrel{!}{=}ET,
\end{equation}
where T is the arrival (or protocol) time,
$\hat{U}(t=T)$ $\!=$ $\!\hat{U}_{\mathrm{T}}$ \cite{Rus14a,Wan14}.
(A ``protocol'' refers to a prescription for the time dependence of ${h}$,
$\hat{H}$, or $\hat{U}$.) In addition to (or instead of) the protocol time $T$,
we may also consider
\begin{equation}
\label{C}
  C\equiv\int_0^T\Bigl\|\frac{\mathrm{d}\hat{H}}{\mathrm{d}t}\Bigr\|\mathrm{d}t
\end{equation}
as a measure of the complexity of the protocol that integrates the changes that
have to be made on $\hat{H}$ via the control fields.

If the optimization problem comprises two objectives such as minimising a
distance $D$ or maximising a fidelity $F$ with minimum amount of (\ref{L}) or
(\ref{C}), then an \emph{approximate} approach consists in first finding an
$h(t)$ that optimizes an objective function $J_1$ under a fixed constraint
$J_2$. Here, $J_1$ represents $D$ or $F$, while $J_2$ may represent
$L$ or $C$. This can be formulated as an Euler-Lagrange equation
\begin{equation}
\label{ELE}
  \frac{\delta{J_1}}{\delta{h}}-\lambda\frac{\delta{J_2}}{\delta{h}}=0,
\end{equation}
where the Lagrange multiplier $\lambda$ must finally be substituted with
the given constant such as $L$ or $C$. This optimisation is then repeated with
stepwise decreased $L$ or $C$, until the deterioration of the achievable $J_1$
exceeds a certain threshold. Equivalently, (\ref{ELE}) may also be thought of
optimizing $J_2$ under the constraint of constant $J_1$. Eq.~(\ref{ELE}), which
contains both derivatives in a symmetric way, merely states the linear
dependence of the functional derivatives at an extremal point ${h}$ in the
function space $\{h\}$.

In the discrete case, the time integrals in Eqs.~(\ref{L}) and (\ref{C}) reduce
to sums over time intervals with constant $\hat{H}$, and in (\ref{C}) we assume
that each jump of $\hat{H}$ gives a fixed finite contribution. To gain some
intuition into the meaning of $C$, we may think of navigating through a
classical rectangular grid. There is a set of shortest paths connecting the
diagonal corners, but they are not equivalent with respect to the number of
turns the navigator has to make along its way. In the quantum context, the
number of switches equals the number of intervals with constant $\hat{H}$, which
may be thought of ``gates''. In contrast to an analytical design of quantum
circuits, the circuit is here generated numerically, however. Since each switch
to a different $\hat{H}$ changes the instantaneous eigenbasis, we may thus think
rather of ``layers'' drawing an analogy to classical artificial neural networks
\cite{bookRojas}.
\begin{acknowledgments}
This work was supported in part by the Austrian Science Fund (FWF) through
project F04012, and by the Templeton World Charity Foundation (TWCF).
\end{acknowledgments}
%
\end{document}